\documentclass[aps,prl,reprint,superscriptaddress,letterpaper,twocolumn,longbibliography,floatfix]{revtex4-2}

\usepackage{polyglossia}
\setmainlanguage{english}
 
\usepackage{amsmath}
\usepackage{amssymb}
\usepackage{amsthm}
\usepackage{amsfonts}
\usepackage{mathtools}
\usepackage{mathrsfs}
\usepackage{dsfont} 
\usepackage{graphicx}
\usepackage{bm}
\usepackage{bbm}
\usepackage{empheq}
\usepackage{cases}
\usepackage{euscript}
\usepackage{xfrac}
\usepackage[shortlabels]{enumitem}

\usepackage{multirow}

\usepackage[svgnames,dvipsnames,x11names]{xcolor}
\usepackage[colorlinks=true,linkcolor=SpringGreen4,citecolor=blue,urlcolor=Magenta3,filecolor=black]{hyperref}
\newcommand{\suppref}[1]{%
  {\hypersetup{linkcolor=black}%
  \ref{#1}}%
}
\usepackage[draft,todonotes={textsize=scriptsize}]{changes}
\definechangesauthor[name={Vishnu Chavva}, color=violet]{VC}
\definechangesauthor[name={Hugo Ribeiro}, color=SpringGreen4]{HR}
\setuptodonotes{inline}

\newcommand{\bra}[1]{\langle #1|}
\newcommand{\ket}[1]{|#1\rangle}
\newcommand{\braket}[2]{\langle #1|#2\rangle}
\newcommand{\ketbra}[2]{\ket{#1}\!\bra{#2}}

\newcommand{\mm}[1]{\mathrm{#1}}
\newcommand{\abs}[1]{\left|#1\right|}

\newcommand{\di}[1]{\mathop{}\!\mathrm{d} #1}

\def \uc{\mathrm{c}}

\def \uI{\mathrm{I}}

\def \rd{\partial}
\def \pv{\mbox{\boldmath$p$}}

\def \hsigma{\hat{\sigma}}

\def \hH{\hat{H}}

\def \hV{\hat{V}}

\def \hT{\hat{T}}
\def \hS{\hat{S}}

\def \hW{\hat{W}}

\def \hP{\hat{P}}
\def \hQ{\hat{Q}}

\def \hPhi{\hat{\Phi}}

\DeclareFontFamily{OT1}{pzc}{}
\DeclareFontShape{OT1}{pzc}{m}{it}{<-> s * [1.10] pzcmi7t}{}
\DeclareMathAlphabet{\mathpzc}{OT1}{pzc}{m}{it}

\DeclareMathOperator{\sech}{sech}

\makeatletter
\newcommand{\npsection}[1]{%
  \par
  \addvspace{2.0ex plus 0.8ex minus 0.2ex}%
  \begingroup
    \centering
    \normalfont\large\bfseries
    #1\par
  \endgroup
  \nobreak
  \vspace{0.8ex plus 0.2ex minus 0.1ex}%
  \@afterheading
}
\makeatother

\usepackage{needspace}

\makeatletter
\newcounter{suppnote}
\renewcommand{\thesuppnote}{\arabic{suppnote}}
\renewcommand{\p@suppnote}{Supplementary Note~}
\newcommand{\suppnote}[1]{
  \par
  \Needspace*{4\baselineskip}
  \refstepcounter{suppnote}
  \addvspace{2.0ex plus 0.8ex minus 0.2ex}
  \begingroup
    \noindent
    \raggedright
    \normalfont\large\bfseries
    Supplementary Note~\thesuppnote: #1\par
  \endgroup
  \nobreak
  \vspace{0.8ex plus 0.2ex minus 0.1ex}%
  \@afterheading
}
\makeatother

\begin{document}

\title{Exceptional-point braiding with native controls}

\author{Vishnu Chavva}
\affiliation{Department of Physics and Applied Physics, University of Massachusetts Lowell, Lowell, MA 01854, USA}

\author{Nero Osmancevic}
\affiliation{Department of Physics and Applied Physics, University of Massachusetts Lowell, Lowell, MA 01854, USA}
    
\author{Hugo Ribeiro}
\affiliation{Department of Physics and Applied Physics, University of Massachusetts Lowell, Lowell, MA 01854, USA}

\begin{abstract}
Exceptional points define branch-exchange state transfers through holomorphic continuation of non-Hermitian eigenmodes, but
realizing these transfers dynamically remains difficult. Slow encircling does not generally transport the full set of
instantaneous eigenstates, while shortcuts to adiabaticity can require controls outside the native experimental manifold. Here, we
introduce a constrained shortcut-to-adiabaticity principle for exceptional-point braiding using native controls. In a dressed
instantaneous-eigenstate frame, the available controls cancel the accessible transition channels locally, while the residual
channels remain active during the evolution but are constrained to have no net accumulated effect over the closed loop. The
protocol therefore targets the endpoint state transfer selected by ideal adiabatic branch exchange, rather than enforcing complete
local cancellation or adiabatic following throughout the trajectory. We demonstrate the construction in a minimal two-mode
non-Hermitian model equivalent, up to a trace shift and a basis convention, to the effective Hamiltonian used in
dissipative-transmon exceptional-point experiments, where detuning and drive amplitude provide real controls and the relative loss
imbalance fixes the non-Hermitian scale. Smooth real waveforms reshape only these controls, reproduce the branch-exchange transfer,
and remain accurate under calibration errors, exceptional-point uncertainty and finite-bandwidth filtering with modest overhead.
\end{abstract}

\maketitle

\npsection{Introduction}

Non-Hermitian systems show how dissipation can reshape spectra, eigenmodes and
dynamics~\cite{Mohammad2019,Kato1976,heiss2012,Moiseyev2011}. Their exceptional points are spectral branch points where
eigenvalues and eigenvectors coalesce, so loops in parameter space can exchange eigenvalue sheets and define eigenstate
braids~\cite{Uzdin2011,Wojcik2022,patil2022,Guria2024}. Such topology has been reconstructed in
optomechanical~\cite{xu2016,patil2022,Guria2024,hoeller2020}, photonic~\cite{Wang2021,Nguyen2023,Klauck2025,Yang2023,Lee2023},
phononic~\cite{Zhang2021,Jin2022,Hu2025,Fang2021,Cheng2022} and nanomechanical platforms~\cite{Wu2025,Wang2024,Zhang2024}. The
next challenge is to use this spectral topology as a dynamical resource.

An exceptional-point loop defines the target topological state transfer through ideal adiabatic branch exchange.  In non-Hermitian
systems, however, slow driving does not generally realize this operation. Gain-loss imbalance and non-adiabatic effects can drive
the physical state away from the holomorphically continued branch, even for slow
loops~\cite{Kato1976,Uzdin2011,hoeller2020,doppler2016,liu2021,abbasi2022,chen2022}. The problem is therefore not only to measure
the braid, but also to implement the corresponding operation dynamically.

Shortcuts to adiabaticity (STAs) offer a route to this goal~\cite{demirplak2003,berry2009,Hatomura2024}, but their implementation
in non-Hermitian systems raises a practical issue because the prescribed Hamiltonian terms may lie outside the experimentally
accessible control manifold.  Counterdiabatic and dressed-state constructions can require auxiliary dissipative controls,
imaginary couplings or complex-valued fields~\cite{demirplak2003,berry2009,baksic2016,ribeiro2017,Erdamar2026,chavva2025_2}. This
is especially restrictive for effective non-Hermitian Hamiltonians. In optomechanical systems, laser power and detuning jointly
modify the effective mode detunings, intermode coupling and damping rates through the optical
susceptibility~\cite{xu2016,patil2022,Guria2024}. In superconducting circuits,
dissipative-transmon experiments provide a particularly direct example. Postselection on a driven three-level transmon realizes an
effective non-Hermitian two-level system whose microwave-induced coupling, detuning and engineered loss set the coherent and
dissipative parameters of the Hamiltonian~\cite{Naghiloo2019,abbasi2022,chen2022}. The native controls are therefore microwave
amplitudes, phases and detunings, together with an engineered loss scale, rather than arbitrary complex matrix elements or
independently programmable auxiliary control channels~\cite{Naghiloo2019,abbasi2022,chen2022,Erdamar2026}. A useful STA must
therefore be designed directly in the real-control manifold of the experiment.

\begin{figure}[t]
    \includegraphics[width=0.99\columnwidth]{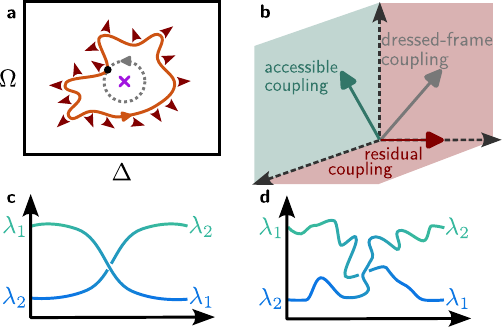}
    \caption{\textbf{Partial local cancellation with a full-cycle flow constraint.}
        \textbf{a,} The dashed grey loop encircles an exceptional point and defines the target endpoint map. The
        constrained-control protocol uses a modified native-control loop, shown in orange, with the same endpoint
        Hamiltonians. Red arrows indicate residual interaction-picture dynamics that remains during the protocol but has zero
        net accumulated effect over the full cycle. \textbf{b,} In the dressed instantaneous-eigenstate frame, the unwanted
        coupling decomposes into an accessible component inside the native-control manifold, shown in green, and an
        inaccessible residual component, shown in red. Only the accessible component can be cancelled locally. \textbf{c,}
        Holomorphic continuation around the bare exceptional-point loop defines the topological braid and hence the endpoint
        state transfer associated with ideal adiabatic branch exchange. \textbf{d,} The modified trajectory does not enforce
        this braid dynamically at intermediate times. Instead, the residual flow is constrained so that the finite-time
        dynamics reproduces the same endpoint state transfer.
        }
    \label{fig:intro}
\end{figure}

This distinction exposes a hierarchy of challenges for near-term implementations. An exceptional-point loop defines a target state
transfer through holomorphic continuation of the eigenmodes, but it does not by itself provide a reliable dynamical protocol.
Slow non-Hermitian encircling does not generally transport the full set of instantaneous eigenstates according to this holomorphic
continuation, because the dynamics tends to select the dynamically dominant branch. Unrestricted counterdiabatic protocols can in
principle restore the desired state transfer, but generally require coupling mechanisms that are not native to the device.
Dressed-state formulations of STAs provide more freedom by changing the frame in which non-adiabatic couplings are cancelled, but
this freedom is still limited by the real-control manifold of the experiment. The available laboratory knobs need not span all
coupling channels between dressed instantaneous eigenstates. The control problem is therefore to identify which couplings can be
cancelled locally with native real controls, and to ensure that the remaining inaccessible channels have no net accumulated effect
over the closed loop.

Here, we construct such STAs by imposing the control constraint from the outset. In a dressed frame, the available controls cancel
the locally accessible part of the dynamics that drives transitions between dressed instantaneous eigenstates. The residual
transition dynamics is not forced to vanish at each instant. Instead, local cancellation expresses the available control fields in
terms of the dressing parameters, whose remaining freedom is then chosen so that the transitions accumulated from the residual
channels vanish over the complete loop~\cite{ribeiro2017,roque2021}. We refer to this principle as partial local cancellation with
a full-cycle flow constraint, illustrated schematically in Fig.~\ref{fig:intro}. The key shift is that the dressed frame is not
used to enforce complete instantaneous transition cancellation. When the experimentally allowed control manifold does not span all
dressed-frame coupling channels, that objective can make the STA construction overconstrained and ill-conditioned near an
exceptional point. We instead use the dressed frame to separate the coupling channels that native controls can cancel locally from
the inaccessible channels whose net effect must vanish over the closed trajectory.

This formulation uses the original exceptional-point loop to define the endpoint state transfer, while the corrected trajectory
need not realize ideal topological transport at intermediate times. We demonstrate the method in a dissipative-transmon-compatible
two-mode model, obtaining smooth native-control waveforms that remain accurate under calibration errors, exceptional-point
uncertainty and finite-bandwidth filtering.

\npsection{Results}

\subsection{Constrained shortcut-to-adiabaticity principle}

We consider an effective non-Hermitian Hamiltonian $\hH[\pv(t)]$ parameterized by real experimental controls $\pv(t)$, such as
detunings, powers or drive amplitudes. These controls define a restricted laboratory manifold in Hamiltonian space, rather than
independent access to arbitrary matrix elements of $\hH$. The map from $\pv(t)$ to $\hH$ may modify coherent and dissipative terms
simultaneously. The control problem is therefore to find a trajectory within this restricted manifold that reproduces the endpoint
state transfer defined by an exceptional-point loop.

The essential step is to relax the requirement of complete instantaneous cancellation in the dressed frame. That requirement is
overconstrained by the native real-control manifold; we instead cancel only the accessible part locally and impose a full-cycle
constraint on the residual interaction-picture flow. Table~\ref{tab:control_hierarchy} summarizes this change relative to
adiabatic encircling, counterdiabatic driving and dressed-state constructions.

\begin{table*}[t]
\centering
\begin{tabular*}{0.99\textwidth}{@{}l@{\extracolsep{\fill}}lll@{}}
\hline
\textbf{Strategy} & \textbf{Dynamical condition} & \textbf{Control assumption} & \textbf{Target} \\
\hline
\multirow{2}{*}{\shortstack[l]{Adiabatic\\encircling}} & Slow driving suppresses & No auxiliary & Branch exchange by \\
& transitions along the EP loop & controls & adiabatic following \\
\hline
\multirow{2}{*}{\shortstack[l]{Counterdiabatic /\\transitionless driving}} & All instantaneous-frame & Auxiliary non-native & Instantaneous \\
& couplings cancelled locally & couplings may be required & branch following \\
\hline
\multirow{2}{*}{\shortstack[l]{Dressed-state\\protocol}} & Relevant dressed-frame & Dressing freedom and controls & Dressed-frame \\
& couplings cancelled locally & enable local cancellation & adiabatic transport \\
\hline
\multirow{2}{*}{\shortstack[l]{Present\\protocol}} & Accessible channel cancelled; & Native real controls only & Endpoint map by \\
& residual flow constrained over loop & & holomorphic continuation \\
\hline
\end{tabular*}
\caption{
\textbf{Control strategies for exceptional-point state transfer.}
The present protocol replaces complete local cancellation with local cancellation of the accessible channel plus a full-cycle
constraint on the residual flow.
}
\label{tab:control_hierarchy}
\end{table*}

The distinction from the dressed-state row is not only a choice of frame or ansatz. With native controls, complete local
cancellation is generally the wrong constrained objective because the laboratory fields access only part of the dressed-frame
transition generator. We therefore modify the STA condition itself:~the controllable part is cancelled locally, while the residual
part is constrained through its full-cycle interaction-picture flow.

The construction starts from a bare loop whose spectrum winds around an exceptional point and hence defines the target state
transfer through ideal adiabatic branch exchange. Because non-Hermitian adiabatic following generally fails, we reshape the loop
using only controls available in the real-control manifold. The remaining dressing freedom is fixed by requiring the residual
interaction-picture flow to have no effect on the target endpoint map.

\subsection{Two-mode state transfer and transmon mapping}

We apply the construction to a minimal two-mode non-Hermitian model directly connected to dissipative-transmon exceptional-point
experiments, where postselection on a driven three-level transmon realizes an effective non-Hermitian two-level system controlled
by microwave drive amplitude, detuning and engineered loss~\cite{Naghiloo2019,abbasi2022,chen2022}. The model therefore represents
the native effective-control problem of an existing quantum exceptional-point platform.

The Hamiltonian has two exceptional points, and the bare loop encircles one of them, defining a branch-exchange state transfer.
Because both instantaneous eigenstates are part of the target map, conventional dressed-state cancellation would require all
relevant dressed-frame couplings to vanish instantaneously. In the present two-mode problem, however, the two laboratory controls
can cancel only one of the two dressed-frame channels locally. Enforcing complete cancellation pushes the remaining condition onto
the dressing parameters, leading to singular or ill-conditioned nonlinear equations near the exceptional point, as shown in
\ref{SupSec:DressedSTA}. The present strategy instead cancels one channel locally and constrains the accumulated residual flow
over the full cycle.

We consider the two-mode non-Hermitian Hamiltonian
\begin{equation}
    \hH_\mm{sym}(t) = -\left[\Delta(t)+i\frac{\Gamma_0}{2}\right]\hat{\sigma}_z + \Omega(t)\hat{\sigma}_x,
    \label{eq:H_sym}
\end{equation}
in which the detuning $\Delta (t)$ and coherent coupling $\Omega (t)$ are tunable real controls, while the relative dissipation
scale $\Gamma_0$ is fixed and cannot be independently modulated. The Hamiltonian has exceptional points at
$(\Delta,\Omega)=(0,\pm\Gamma_0/2)$. 

For comparison with the dissipative-transmon realization, the postselected transmon Hamiltonian can be written as
\begin{equation}
    \begin{aligned}
        \hH_\mm{tr} &= \Delta_\mm{tr} \ketbra{e}{e} - \frac{i}{2} \left( \gamma_e \ketbra{e}{e} + \gamma_f \ketbra{f}{f} \right) \\
        &\phantom{={}} + J \left(\ketbra{f}{e} + \ketbra{e}{f}\right),
    \end{aligned}
    \label{eq:Htransmon}
\end{equation}
where $J$ is the microwave-drive-induced coupling, $\Delta_\mm{tr}$ is the transmon detuning, and $\gamma_e$ and $\gamma_f$ are
the decay rates of the two states forming the postselected non-Hermitian manifold~\cite{Naghiloo2019,abbasi2022,chen2022}.
Removing an irrelevant scalar trace gives
\begin{equation}
    \hH_\mm{tr}^{(0)} = \frac{1}{2} \left(\Delta_\mm{tr} -i \frac{\gamma_e -\gamma_f}{2}\right) \hat{\sigma}_z + J \hat{\sigma}_x,
    \label{eq:Htransmon_tracefree}
\end{equation}
with $\hat{\sigma}_z = \ketbra{e}{e} - \ketbra{f}{f}$ and $\hat{\sigma}_x = \ketbra{f}{e} +  \ketbra{e}{f}$.
Equation~\eqref{eq:H_sym} is therefore obtained, up to this basis convention, through the identification $\Omega=J$,
$\Delta=-\Delta_{\rm tr}/2$ and $\Gamma_0=(\gamma_e-\gamma_f)/2$. Thus, $\Gamma_0$ is set by the relative loss imbalance after
removal of the common trace loss.

Using reported dissipative-transmon decay rates~\cite{Naghiloo2019,abbasi2022}, this mapping gives $\Gamma_0$ of order a few
inverse microseconds and $J_\mm{EP} = \Gamma_0 / 2$ of order $1~\mm{rad}/\mu \mm{s}$. The loop durations used below, $\Gamma_0 t_0
= 5$--$10$, therefore correspond to microsecond-scale protocols, comparable to existing transmon encircling experiments. Detailed
estimates are given in Methods. The loss imbalance sets the non-Hermitian scale, but is not used as an independently modulated
shortcut control.

We choose the bare loop to wind once around one of them,
\begin{equation}
    \begin{aligned}
        \Delta (t) &= \Delta_0 \sin\left( 2\pi s(t) + \varphi \right), \\
        \Omega (t) &= -\Omega_0 - \Delta_0 \cos\left( 2\pi s(t) + \varphi \right),
    \end{aligned}
    \label{eq:2ModeContour}
\end{equation}
where $\varphi$ sets the base point. The STA is constrained to the same real-control manifold:~it can only
reshape the detuning and coherent coupling, $\Delta(t) \to \Delta (t) +g_z (t)$ and $\Omega (t) \to \Omega (t) +g_x (t)$, while
leaving $\Gamma_0$ unchanged. Thus, the protocol modifies only the native real controls of the model.

We parameterize the dressing transformation by two time-dependent angles, whose explicit form is given in Methods. These dressing
angles are written as smooth finite Fourier series, and the Fourier coefficients provide the free parameters of the construction.
For any candidate set of coefficients, the dressing defines a frame in which the transition-generating dynamics is split into an
accessible part and a residual part. Local cancellation of the accessible part algebraically determines the corresponding real
fields $g_x (t)$ and $g_z (t)$, so these control corrections become functions of the Fourier coefficients. The coefficients are
then chosen so that the residual interaction has no net effect over the full cycle. In the two-mode model, the residual channel
has a simple structure and the full-cycle condition reduces to a scalar nonlinear equation, which we solve by minimizing its
squared residual as a stable root-finding strategy.

\subsubsection{Reproducing the endpoint state transfer with real controls}

We first consider the ideal calibrated model, where the computed real-control waveforms are implemented exactly. We evaluate the
dynamics in the adiabatic frame built from instantaneous eigenstates followed by holomorphic continuation around the exceptional
point. In this frame, $P_{i,j}(t)$ is the normalized transition probability from the initial continued branch $i$ to the continued
branch $j$. Successful realization of the branch-exchange state transfer therefore corresponds to $P_{i,i}(t_0)=1$, since the
branch label is preserved in the holomorphically continued frame even though the associated instantaneous eigenstate of the bare
Hamiltonian is exchanged after the loop. We denote the corresponding state-transfer error by $\varepsilon_{i,j}(t)=1-P_{i,j}(t)$.
Thus, the final error tests whether the real-control construction reproduces the state transfer selected by ideal adiabatic branch
exchange, rather than whether the state follows an instantaneous eigenstate at every intermediate time.

\begin{figure}[t]
    \includegraphics[width=0.99\columnwidth]{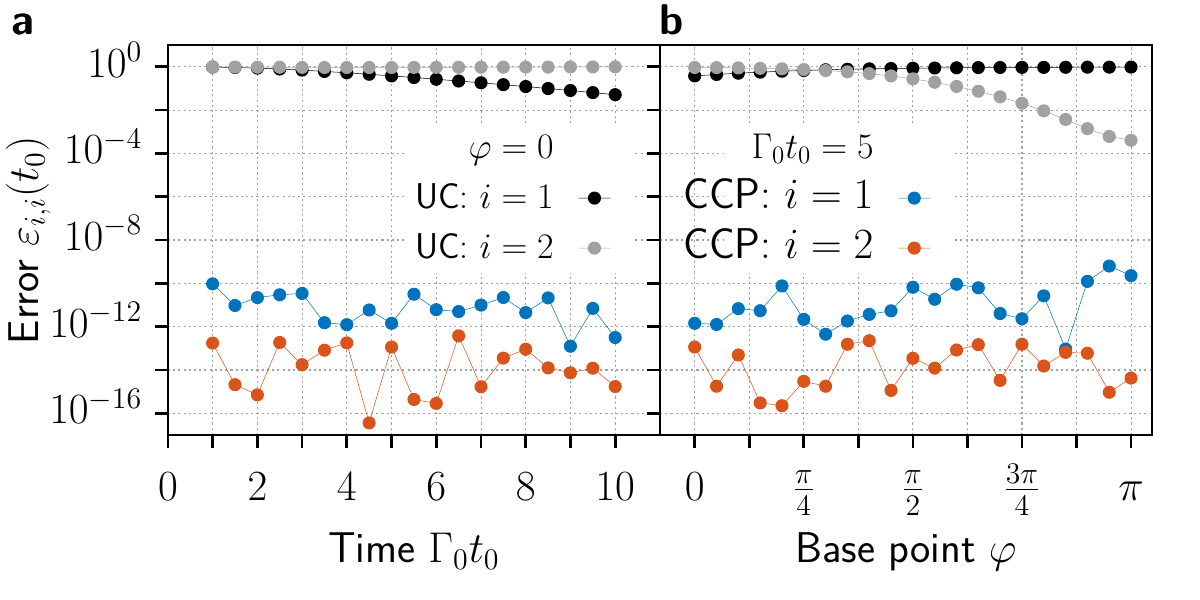}
    \caption{\textbf{Performance across loop duration and base point.}
        \textbf{a,} Final state-transfer error $\varepsilon_{i,i}(t_0)$ as a function of loop duration for the two initial
        instantaneous eigenstates. Grey and black points show uncontrolled dynamics; blue and orange points show the
        constrained-control protocol. \textbf{b,} Final state-transfer error as a function of the loop base point $\varphi$ for
        fixed loop duration $(\Gamma_0 t_0 =5)$. Low state-transfer error is maintained across base points, demonstrating that the
        protocol does not rely on a specially chosen starting point on the loop.
    }
    \label{fig:phi_comparison}
\end{figure}

Figure~\ref{fig:phi_comparison}a shows the final state-transfer error $\varepsilon_{i,i} (t_0)$ as a function
of loop duration. Without control, the dynamics does not reliably produce the target operation, reflecting the failure of ordinary
non-Hermitian adiabatic following.  With partial local cancellation and full-cycle flow constraint, the error is strongly
suppressed. This low error is not imposed as the defining constraint; it is an independent diagnostic that the resulting
real-control trajectory reproduces the target endpoint state transfer.

Figure~\ref{fig:phi_comparison}b shows that for the model and parameter range examined here, successful transfer does not require
a uniquely selected base point of the exceptional-point loop. This differs from recent approaches that restore adiabatic state
transfer through special time-modulated trajectories, including STA extensions and experimental photonic
implementations~\cite{Arkhipov2024,wu2024,Wang2026}. Here, accurate endpoint transfer does not require a specially chosen base
point. The accessible dressed-frame channel is cancelled locally and the residual flow is constrained over the full cycle,
allowing the protocol to remain accurate as the starting point is varied. \ref{SupSec:TwoModeAlternate} shows that the
construction is also not tied to the particular choice of locally cancelled dressed-frame coupling channel; one may cancel the
opposite channel locally and constrain the remaining residual flow over the full cycle. 

\begin{figure}[t]
    \includegraphics[width=0.99\columnwidth]{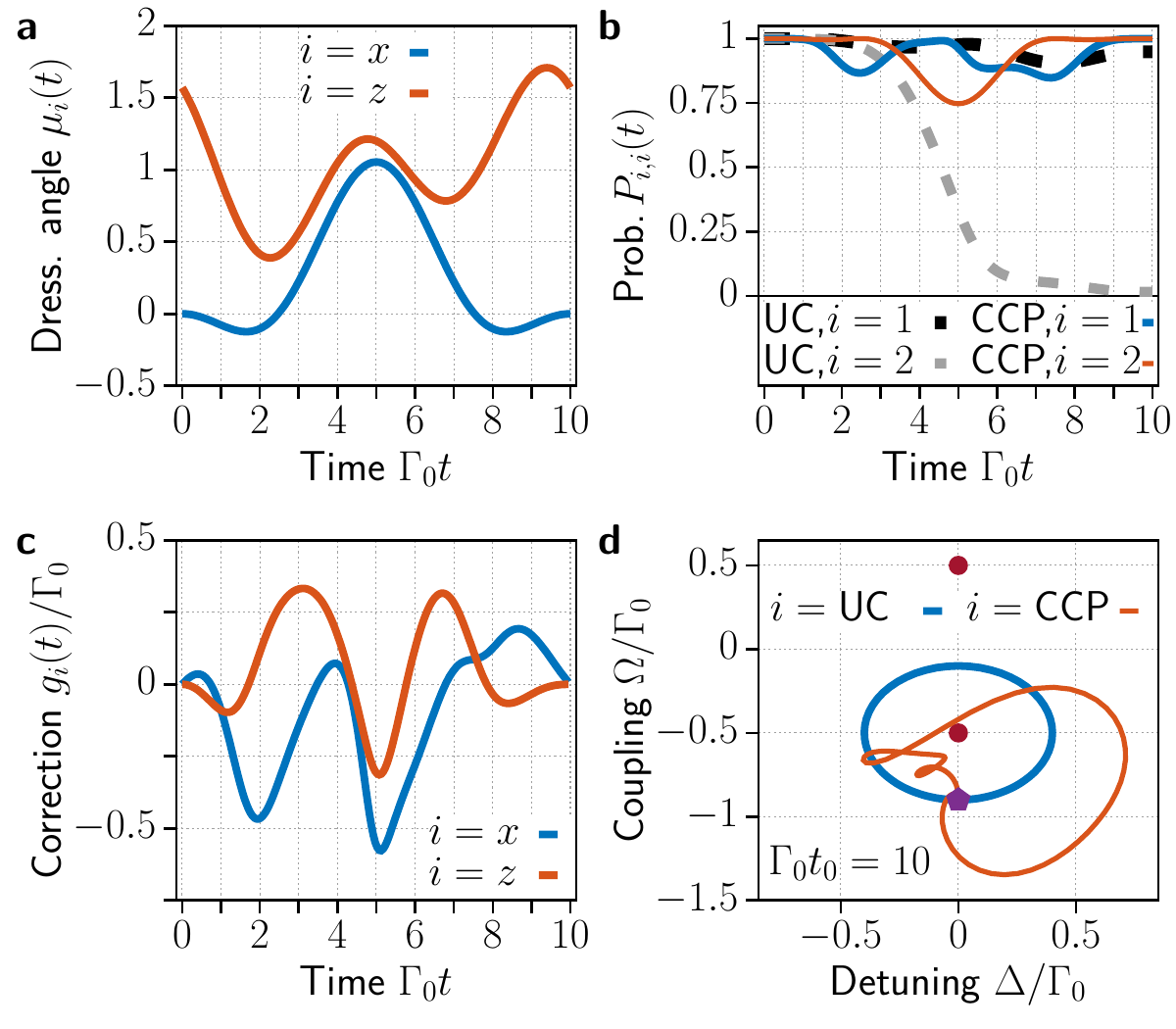}
    \caption{\textbf{Real control waveforms and constrained loop dynamics.}
        \textbf{a,} Numerically determined dressing angles $\mu_x(t)$ and $\mu_z(t)$ satisfying the full-cycle flow constraint for
        $\Gamma_0 t_0=10$. \textbf{b,} Instantaneous-eigenstate populations during the loop for uncontrolled and
        constrained-control dynamics. \textbf{c,} Real laboratory-frame corrections $g_x(t)$ and $g_z(t)$. \textbf{d,} Bare and
        modified trajectories in the $(\Delta,\Omega)$ plane, with exceptional points marked by red dots and the loop base point
        by a purple pentagon. The modified trajectory remains within the assumed real-control manifold. Because the correction
        fields vanish at the endpoints, the bare and controlled Hamiltonians coincide there and share the same initial and final
        instantaneous eigenstates.
    }
    \label{fig:sample_result}
\end{figure}

Figure~\ref{fig:sample_result} shows a representative protocol. The dressing functions generate the real corrections $g_x(t)$ and
$g_z(t)$, and the controlled dynamics maps each initial branch to the state selected by ideal adiabatic branch exchange. Because
the correction fields vanish at $t=0$ and $t=t_0$, the controlled and bare Hamiltonians coincide at the endpoints and share the
same initial and final instantaneous eigenstates. This physical endpoint matching does not require the auxiliary dressing
transformation to reduce to the identity. Its boundary conditions instead ensure that the initial transformation into the dressed
frame is reversed by the final transformation back into the instantaneous-eigenstate frame without altering the target endpoint
flow.

These ideal-theory results establish the main point: smooth real waveforms in the native control manifold can reproduce the state
transfer selected by adiabatic branch exchange without auxiliary dissipation engineering, imaginary-valued fields or additional
coupling channels. The protocol therefore probes the endpoint state transfer associated with the original exceptional-point loop,
without requiring the finite-time controlled trajectory to retain the topological robustness of ideal adiabatic following.

\subsubsection{Robustness and control overhead}

\begin{figure*}[t]
    \includegraphics[width=1.99\columnwidth]{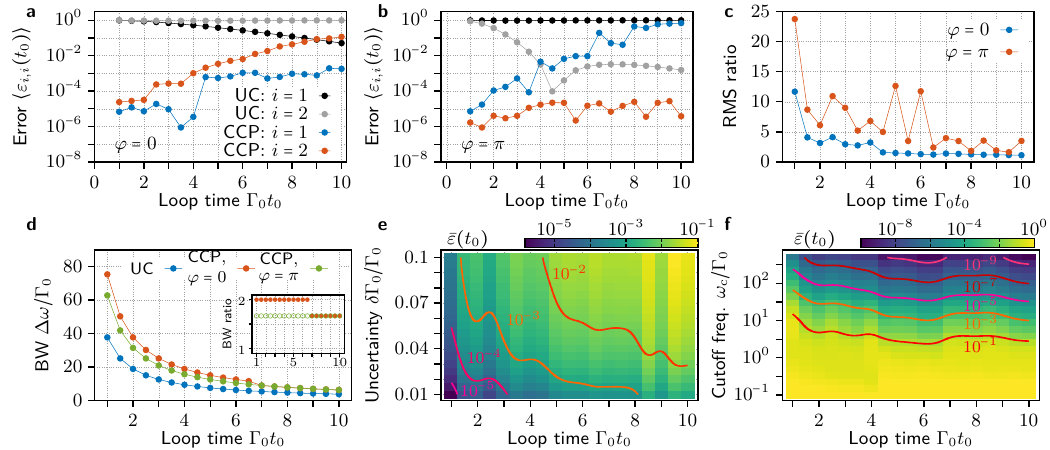}
    \caption{
        \textbf{Robustness and control overhead.}
        \textbf{a,b,} Final state-transfer error $\langle\varepsilon_{i,i}(t_0)\rangle$ averaged over static detuning offsets for
        two loop base points. \textbf{c,} RMS-amplitude overhead of the constrained-control waveform relative to the uncorrected
        loop. \textbf{d,} Absolute 99\% bandwidth and bandwidth overhead. \textbf{e,} Branch-averaged final state-transfer error
        $\bar{\varepsilon} (t_0) = [\varepsilon_{1,1} (t_0) + \varepsilon_{2,2} (t_0)]/2$
        after propagating with a shifted value of $\Gamma_0$, modelling uncertainty in the inferred exceptional-point position.
        \textbf{f,} The same branch-averaged error for an independent finite-bandwidth test, where only the target controls are
        low-pass filtered with cutoff $\omega_\uc$ and $\Gamma_0$ is not shifted.
    }
    \label{fig:robustness}
\end{figure*}

We next test robustness to static detuning offsets, exceptional-point uncertainty, finite control bandwidth and finite pulse
amplitude. For static offsets, the controls are kept fixed and the implemented loop is perturbed without recomputing the dressing
angles. Figure~\ref{fig:robustness}a,b shows that the controlled dynamics remains accurate at experimentally relevant short loop
times and continues to outperform the uncontrolled dynamics over the range shown. The comparison between the two base points shows
that this conclusion does not rely on a specially chosen launch point. Even in the presence of static detuning imprecision, the
controlled protocol remains accurate over the operating range shown, consistent with the full-cycle constraint on the residual
dressed-frame flow.

Figure~\ref{fig:robustness}c,d quantifies the resources required by the modified pulses. In the robust operating regime, the
RMS-amplitude overhead is typically $R_\mm{rms} \sim 1$--$2.5$, and the bandwidth overhead is only $R_\mm{bw} \sim 5/3$--$2$.
Figure~\ref{fig:robustness}e,f shows the branch-averaged state-transfer error, $\bar{\varepsilon} (t_0) = [\varepsilon_{1,1} (t_0)
+ \varepsilon_{2,2}(t_0)]/2$ in two independent tests: uncertainty in $\Gamma_0$ and low-pass filtering, respectively.  This
simple average is used only to display the robustness of both transported branches on a single color scale; provided the cutoff
retains the relevant spectral content, the protocol remains accurate over the operating region shown.  Within the parameter ranges
shown, these checks show that low final state-transfer error is obtained in the same regimes where the required amplitude and
bandwidth overhead remain modest.

This resource analysis connects the full-cycle constraint on the residual flow to laboratory waveform constraints. A control
protocol that reproduces the endpoint state transfer only by using very large amplitudes or high-frequency features would be of
limited experimental value. The operating regions identified here avoid this regime, with the corrected waveforms remaining
smooth, finite-bandwidth modifications of the original loop.

The overheads in Fig.~\ref{fig:robustness} have a direct interpretation for the dissipative-transmon realization discussed above.
Combining the transmon scales estimated in Methods with the overheads in Fig.~\ref{fig:robustness}c,d shows that the corrected
effective waveforms remain within the same order of magnitude as reported coherent-control amplitudes and bandwidths. The
corrected effective waveforms $\Omega(t)+g_x(t)$ and $\Delta(t)+g_z(t)$ correspond, through the calibrated transmon control map,
to shaped microwave-amplitude and detuning pulses in the native coherent-control manifold. The resource estimates therefore
quantify the extra coherent-control amplitude and bandwidth required at the effective-Hamiltonian level, with no additional
control over the dissipative channel. This contrasts with counterdiabatic implementations, which enforce instantaneous eigenstate
following by introducing dedicated auxiliary controls~\cite{Erdamar2026}. Here, no additional control axis is introduced. The
protocol remains within the native coherent-control manifold of the dissipative-transmon Hamiltonian, reshaping only the microwave
drive and detuning, while the loss channel sets the non-Hermitian scale rather than serving as an independently modulated control.

More generally, the method uses a calibrated real-control manifold to identify the accessible dressed-frame channels and constrain
the residual flow. Dissipative transmons provide the most direct near-term test, while optomechanical systems provide a
complementary setting in which laser power and detuning modify effective detuning, coupling and damping through a device-specific
optical susceptibility~\cite{xu2016,patil2022,Guria2024}.

Taken together, the ideal-theory prediction, robustness checks and pulse-cost analysis show that the method produces real, smooth
and finite-bandwidth waveforms inside the native control manifold, while retaining low final state-transfer error for the endpoint
state transfer selected by exceptional-point branch exchange.

\npsection{Discussion}

The protocol introduced here separates the topological definition of an exceptional-point state transfer from the constrained
dynamics used to realize it. The exceptional-point loop supplies an endpoint map through holomorphic continuation of the
eigenmodes, defining the state transfer associated with ideal adiabatic branch exchange. Because the added controls vanish at the
endpoints, the controlled and bare Hamiltonians share the same initial and final instantaneous eigenstates. The boundary
conditions on the auxiliary dressing transformation do not require it to reduce to the identity at the endpoints. They require the
initial transformation into the dressed frame and the final transformation back into the instantaneous-eigenstate frame to
preserve the target subspace and to reverse one another without altering the desired projected flow. The protocol therefore
realizes the abstract endpoint map as a physical state transfer between the instantaneous eigenstates of the original
exceptional-point loop, enabling state-transport phenomena associated with exceptional-point topology to be probed directly.

A conventional STA construction would try to realize this map by cancelling all transition-generating couplings, either in the
instantaneous-eigenstate frame or in a suitably dressed frame. Under native real-control constraints, however, the available
Hamiltonian parameters need not provide enough independent knobs to cancel all dressed-frame couplings locally. This mismatch is
especially important for effective non-Hermitian Hamiltonians, where laboratory knobs can modify coherent and dissipative matrix
elements in a correlated way. Our construction addresses this constrained regime by changing the STA condition itself, rather than
trying to solve an overconstrained complete-cancellation problem.

The closest experimental realization is the dissipative-transmon platform used in earlier exceptional-point experiments. Because
the trace-shifted postselected Hamiltonian maps onto Eq.~\eqref{eq:H_sym}, the protocol uses the same native coherent controls
while leaving the loss channel to set the non-Hermitian scale, without adding an independently modulated dissipative control
channel~\cite{Erdamar2026}.

The same constrained-control principle should also be useful in optomechanical systems, where exceptional-point topology and
braiding have been extensively explored~\cite{xu2016,patil2022,Guria2024}. In that setting, however, laboratory controls generally
enter through a device-specific optical susceptibility, so the effective detuning, coupling and damping parameters are not
independently tunable. The present formulation is designed for precisely this situation:~one first identifies the calibrated
real-control manifold and then separates the dressed-frame dynamics into controllable and residual channels. The two-mode model
studied here provides the minimal experimentally relevant realization in which local cancellation of one dressed-frame channel
leaves a single residual transition direction, so the residual-flow constraint reduces to a scalar condition. In larger systems,
the same separation generally leads to matrix-valued full-cycle constraints on the residual interaction-picture flow, which can be
treated with the constrained root-finding or continuation strategies described above. In this sense, dissipative transmons provide
the most direct near-term test of the protocol, while optomechanics illustrates why a control-manifold formulation is needed
across non-Hermitian platforms. The modest amplitude and bandwidth overheads found here suggest a practical route from
spectroscopic exceptional-point topology to experimentally implementable non-Hermitian state-transfer protocols.

\npsection{Methods}

\subsection{General coherent-control construction}

We consider an effective non-Hermitian Hamiltonian $\hH[\pv(t)]$ parameterized by real experimental controls $\pv(t)$, such as
detunings, powers or drive amplitudes. These controls define a restricted laboratory manifold in Hamiltonian space, rather than
independent access to arbitrary matrix elements of $\hH$. The map from $\pv(t)$ to $\hH$ may modify coherent and dissipative terms
simultaneously. The control problem is therefore to find a trajectory within this restricted manifold that reproduces the endpoint
state transfer defined by an exceptional-point loop.

Let $\ket{\psi_j (t) }$ and $\bra{\psi_j (t)}$ be the right and left instantaneous eigenstates of $\hH[\pv (t)]$, with
eigenvalue $\lambda_j (t)$ and biorthogonal normalization $\braket{\psi_j (t)}{\psi_i (t)} = \delta_{i,j}$. The transformation to
the instantaneous-eigenstate frame is denoted $\hS_\mm{ad} (t)$. In this frame, the Hamiltonian separates into
\begin{equation}
    \begin{aligned}
        \hH_\mm{ad} (t) &= \hS_\mm{ad}^{-1} (t) \hH[\pv (t)] \hS_\mm{ad} (t) - i \hS_\mm{ad}^{-1} (t) \rd_t \hS_\mm{ad} (t) \\
        &= \hH_{\mm{ad},0} (t) + \hV_\mm{nad} (t).
    \end{aligned}
    \label{eq:Had}
\end{equation}
Here, $\hH_{\mm{ad},0} (t) = \sum_j (\lambda_j (t) -i \braket{\psi_j (t)}{\rd_t \psi_j (t)})  \ketbra{\psi_j}{\psi_j}$ is
diagonal, whereas $\hV_\mm{nad} (t)$ contains the non-adiabatic couplings generated by the time dependence of $\hS_\mm{ad}(t)$.
The construction of $\hS_\mm{ad} (t)$, including the holomorphic continuation of the eigenvalues and eigenmodes around the
exceptional point, is described in \ref{SupSec:ChangeOfFrame}.

A laboratory-frame control Hamiltonian $\hW(t)$ is then added, with $\hW(t)$ restricted to the experimentally available coherent
operator set. In the instantaneous-eigenstate frame this gives $\hW_\mm{ad}(t)=\hS_\mm{ad}^{-1}(t)\hW(t)\hS_\mm{ad}(t)$. We denote
the modified Hamiltonian expressed in the instantaneous-eigenstate frame of the bare Hamiltonian by
\begin{equation} 
    \hH_\mm{mod,ad}(t) = \hH_\mm{ad}(t) + \hW_\mm{ad}(t). 
    \label{eq:HmodAd} 
\end{equation} 
We introduce a dressing transformation $\hS_\mm{dr} (t)$ and define the dressed-frame Hamiltonian
\begin{equation}
    \begin{aligned}
        \hH_\mm{dr} (t) &= \hS_\mm{dr}^{-1} (t) \hH_\mm{mod,ad} (t) \hS_\mm{dr} (t) - i \hS_\mm{dr}^{-1} (t) \rd_t \hS_\mm{dr} (t)\\
        &= \hH_0 (t) + \hV (t).
    \end{aligned}
    \label{eq:Hdr}
\end{equation}
The dressed-frame Hamiltonian is decomposed into an ideal part $\hH_0(t)$, which generates the desired evolution, and a dressed
non-adiabatic coupling term $\hV(t)$, which contains the couplings between dressed instantaneous eigenstates. We denote the flows
generated by $\hH_\mm{mod,ad}(t)$ and $\hH_\mm{dr}(t)$ by $\hPhi_\mm{mod,ad}(t)$ and $\hPhi_\mm{dr}(t)$, respectively.

The coherent-control procedure implements partial local cancellation with a full-cycle flow constraint by partitioning $\hV(t)$
according to the available controls.  Let $l$ denote the set of dressed-frame coupling channels accessible to the native
coherent controls. These channels are cancelled locally by imposing
\begin{equation}
    \bra{\psi_{\mm{dr},j}} \hV (t) \ket{\psi_{\mm{dr},i}} =0, \quad \forall i,j  \in l,
    \label{eq:CondW}
\end{equation}
with $\ket{\psi_{\mm{dr},i}} = \hS_\mm{dr}^{-1} (t) \ket{\psi_i (t) }$ denoting a dressed instantaneous eigenstate. The remaining
channels, denoted $\bar{l}$, are not required to vanish at each instant. Instead, the protocol is constructed so that the
interaction-picture flow has no net effect over the complete loop. This full-cycle condition can be formulated by
first isolating the residual dressed-frame coupling Hamiltonian,
\begin{equation}
    \hV_{\bar{l}} (t) = \sum_{i,j \in \bar{l}} \bra{\psi_{\mm{dr},j}} \hV (t) \ket{\psi_{\mm{dr},i}} \ketbra{\psi_{\mm{dr},j}}{\psi_{\mm{dr},i}},
    \label{eq:VRemaining}
\end{equation}
and the dressed-frame flow operator generated by the ideal dynamics,
\begin{equation}
    \hPhi_0 (t) = \hT \exp\left[-i \int_0^t \di{t_1} \hH_0 (t_1) \right].
    \label{eq:Phi0}
\end{equation}
The residual interaction-picture Hamiltonian is then
\begin{equation}
    \hV_\uI (t) = \hPhi_0^{-1} (t) \hV_{\bar{l}} (t) \hPhi_0 (t).
    \label{eq:VI}
\end{equation}
The full-cycle flow constraint is satisfied when the residual dressed-frame flow over one loop is trivial,
\begin{equation}
    \hPhi_\uI (t_0) = \hT \exp\left[-i \int_0^{t_0} \di{t} \hV_\uI (t) \right] = \mathbbm{1},
    \label{eq:IntPictFlow}
\end{equation}
or, more generally, when $\hPhi_\uI(t_0)$ is diagonal and contributes only phases or normalization factors that do not change the
target endpoint map. Thus, the method does not require the available control Hamiltonian to cancel all dressed-frame couplings
instantaneously. Instead, the dressing and the native controls are determined self-consistently. The available control Hamiltonian
$\hW(t)$ cancels the accessible coupling channels locally, while the dressing freedom is used to constrain the residual channels
to generate a trivial, diagonal, or otherwise irrelevant accumulated flow over the full cycle.

In practice, the construction proceeds in three steps. First, one chooses a bare loop $\pv(t)$ that winds around the target
exceptional point or exceptional-point manifold. Second, one chooses boundary conditions on the dressing transformation so that
the initial transformation into the dressed frame and the final transformation back into the instantaneous-eigenstate frame
preserve the target subspace and do not alter the desired endpoint flow. Let $\hP$ project onto the subspace
of states to be transported and let $\hQ=\mathbbm{1}-\hP$. A sufficient set of boundary conditions is
\begin{equation} 
    \begin{aligned} 
        &\text{(i)}\quad &&\left[ \hP\hS_\mm{dr}(t_0)\hP,\, \hP\hPhi_\mm{dr}(t_0)\hP \right] = \mathbf{0}, \\ 
        &\phantom{\text{(i)}} \text{or} \quad &&\left[\hP\hS_\mm{dr}^{-1}(0)\hP,\, \hP\hPhi_\mm{dr}(t_0)\hP \right] = \mathbf{0}, \\ 
        &\text{(ii)}\quad &&\hP \hS_\mm{dr}(t_0) \hS_\mm{dr}^{-1}(0) \hP = \hP, \\ 
        &\text{(iii)}\quad &&\left[ \hP,\hS_\mm{dr}^{-1} (0) \right] = \left[ \hP,\hS_\mm{dr}(t_0) \right] = \mathbf{0}. 
    \end{aligned} 
\label{eq:DressingBoundaryConditions} 
\end{equation}
The first condition ensures that the dressing transformation at the boundaries does not change the state transfer generated by the
dressed-frame endpoint flow. The second ensures that the final transformation back to the instantaneous-eigenstate frame reverses
the initial transformation into the dressed frame within the target subspace. The third ensures that both transformations preserve
the target subspace and do not mix it with its complement. Together, these conditions permit a nontrivial dressing transformation
at the protocol boundaries without altering the desired projected endpoint flow. Their derivation is given under ``Boundary
conditions for the dressing transformation'' in the Methods.

Third, for each candidate dressing transformation, the available coherent controls are chosen to cancel the channels in
$l$ locally. The remaining free parameters of $\hS_\mm{dr}(t)$ are determined by the full-cycle constraint on the
residual flow in Eq.~\eqref{eq:IntPictFlow}. This condition generally gives nonlinear coupled equations for the dressing
parameters. They can be solved directly, by continuation from nearby solutions or with constrained numerical methods. A
minimization of the residual-flow norm may be used as a root-finding method, but is not the defining control objective.

We give a generalization of Eq.~\eqref{eq:IntPictFlow} in \ref{SupSec:RelaxedFlowConstraint} to construct protocols that only
transport a subset of modes. 

\subsection{Two-mode non-Hermitian model}

The two-mode model used in the main text is defined in Eq.~\eqref{eq:H_sym}. With our convention, the exceptional points occur at
$(\Delta,\Omega)=(0,\pm\Gamma_0/2)$. The bare loop is given by Eq.~\eqref{eq:2ModeContour}, where $\Delta_0$ is the loop radius,
$\Omega_0$ is the loop center, $\varphi$ is the base point, and $s (t)$ is a smooth ramp. We use
\begin{equation}
    s (t) = d \left[6\left(\frac{t}{t_0}\right)^5  -15\left( \frac{t}{t_0} \right)^4+10\left( \frac{t}{t_0}
    \right)^3\right],
    \label{eq:SmoothRamp}
\end{equation}
where $d=\pm 1$ sets the loop orientation. This ramp satisfies $\dot{s} (0)=\dot{s}(t_0) =
\ddot{s} (0) = \ddot{s} (t_0) = 0$, ensuring smooth switch-on and switch-off of the parameter loop.

The instantaneous-eigenstate frame is constructed from the biorthogonal eigenstates of $\hH_\mm{sym} (t)$. Because the loop winds
around an exceptional point, the eigenvalues and eigenvectors are followed by holomorphic continuation so that the two Riemann
sheets are tracked consistently throughout the cycle. The explicit form of the adiabatic-frame transformation is given in
\ref{SupSec:ChangeOfFrame}.

\subsection{Connection to dissipative-transmon scales}

The postselected dissipative-transmon Hamiltonian maps onto Eq.~\eqref{eq:H_sym} after subtracting the scalar trace and choosing
the Pauli convention described in the main text. If the two states in the postselected manifold have decay rates $\gamma_e$ and
$\gamma_f$, the trace-subtracted non-Hermitian scale is set by the relative loss imbalance,
\begin{equation}
    \Gamma_0 = \frac{\gamma_e - \gamma_f}{2}.
    \label{eq:Gamma0Transmon}
\end{equation}
The exceptional point of Eq.~\eqref{eq:H_sym}, located at $\Omega = \Gamma_0/2$, therefore corresponds to $J_\mm{EP} = \Gamma_0/2
= (\gamma_e-\gamma_f)/4$ in the dissipative-transmon notation.

For the earlier dissipative-transmon tomography experiment~\cite{Naghiloo2019}, the reported rates $\gamma_e = 6.7~\mu \mm{s}^{-1}$
and $\gamma_f = 0.25~\mu \mm{s}^{-1}$ give $\Gamma_0 =3.23~\mu \mm{s}^{-1}$ and $J_\mm{EP} = 1.61~\mm{rad}/\mu \mm{s}$, consistent
with the fitted exceptional-point threshold $J_0 = 1.71 \pm 0.07~\mu \mm{s}^{-1}$~\cite{Naghiloo2019}. For the dissipative-transmon
state-control experiment~\cite{abbasi2022}, the reported rates $\gamma_e = 6.2~\mu \mm{s}^{-1}$ and $\gamma_f = 0.32~\mu \mm{s}^{-1}$
give $\Gamma_0 \simeq 2.94~\mu \mm{s}^{-1}$ and $J_\mm{EP} \simeq 1.47~\mm{rad}/\mu \mm{s}$~\cite{abbasi2022}. The dimensionless
loop durations used in the main text, $\Gamma_0 t_0 = 5$--$10$, therefore correspond to $t_0 \simeq 1.7$--$3.4~\mu \mm{s}$ for these
transmon parameters.

The same state-control experiment used microsecond-scale parameter cycles, with reported loop periods $T = 800~\mm{ns}$ and $T =
1.5~\mu \mm{s}$, and coherent control scales $J_\mm{max} = 30~\mm{rad}/\mu \mm{s}$ and $\Delta_\mm{max} = 10\pi~\mm{rad}/\mu
\mm{s}$~\cite{abbasi2022}. These values place the effective timescales and coherent-control coordinates of Eq.~\eqref{eq:H_sym} in
the regime of existing dissipative-transmon exceptional-point experiments. These estimates should be compared with the overheads
extracted from Fig.~\ref{fig:robustness}c,d. In the robust operating region, the constrained-control protocol increases the
effective RMS amplitude by a factor $R_\mm{rms} \sim 1$--$2.5$ and the required bandwidth by $R_\mm{bw} \sim 5/3$--$2$. Thus,
relative to the dissipative-transmon scales quoted above, the corrected waveforms require only order-unity increases in
coherent-control amplitude and bandwidth.

In the interference experiment, the additional $\ket{h}$ level is used as a phase reference and is included in the full Lindblad
modelling together with state-dependent dephasing rates $\gamma_{2e}$, $\gamma_{2f}$, and $\gamma_{2h}$. The reported values
include the reference-level loss rate $\gamma_h = 0.36~\mu \mm{s}^{-1}$ and the dephasing rates $\gamma_{2e} = 3.7~\mu
\mm{s}^{-1}$, $\gamma_{2f} = 0.9~\mu \mm{s}^{-1}$ and $\gamma_{2h} =1.4~\mu \mm{s}^{-1}$~\cite{abbasi2022}. These additional rates
enter the modelling of the interference signal and its reconstruction, but they do not set the two-level exceptional-point scale.
That scale is fixed by the relative $\ket{e}$--$\ket{f}$ population-loss imbalance in Eq.~\eqref{eq:Gamma0Transmon}.

\subsection{Real coherent controls}

The real coherent-control Hamiltonian is
\begin{equation}
    \hW (t) = g_z(t) \hat{\sigma}_z +g_x (t) \hat{\sigma}_x,
    \label{eq:Wsym}
\end{equation}
so that the modified controls are $\Delta_\mm{mod} (t) = \Delta (t) + g_z (t)$ and $\Omega_\mm{mod} (t) = \Omega (t) +g_x (t)$.
The relative dissipation $\Gamma_0$ is not modified. This real-control Hamiltonian excludes auxiliary dissipative terms and
additional non-Hermitian coupling channels.

Although the laboratory-frame fields $g_x(t)$ and $g_z(t)$ are real, the control Hamiltonian generally has complex matrix elements
in the instantaneous-eigenstate frame. These matrix elements cannot be chosen independently because they must arise from the real
laboratory-frame Hamiltonian in Eq.~\eqref{eq:Wsym}. This constraint imposes algebraic relations between their real and imaginary
parts. For a given dressing transformation, those relations determine which dressed-frame coupling channel can be cancelled
locally by the available controls. The explicit algebraic relations are given in \ref{SupSec:TwoMode}.

\subsection{Two-mode implementation of the coherent-control procedure}

For any choice of dressing transformation, the dressed non-adiabatic coupling term can be written as
\begin{equation}
    \hV (t) = \beta_+ (t) \hat{\sigma}_+ + \beta_- (t) \hat{\sigma}_-
    \label{eq:Vdr2Mode}
\end{equation}
with $\hat{\sigma}_\pm = \frac{1}{2} (\hat{\sigma}_x \pm i\hat{\sigma}_y)$. For a fixed choice of dressing angles, the real coherent
fields $g_x (t)$ and $g_z (t)$ are determined by imposing local cancellation of the accessible channel. In the implementation used
here, this condition is
\begin{equation}
    \beta_+ (t) = 0,
    \label{eq:2ModeCondW}
\end{equation}
which leaves the residual channel
\begin{equation}
    \hV_{\bar{l}} (t) = \beta_- (t) \hat{\sigma}_-.
    \label{eq:2ModeVdr2}
\end{equation}
The residual channel is not required to vanish at each instant. Instead, the dressing angles are chosen so that its
interaction-picture flow has no net effect over the full cycle. Explicit expressions for $\beta_+ (t)$, $\beta_- (t)$, as well as
the form of $g_x (t)$ and $g_z (t)$ leading to $\beta_+ (t) = 0$, are given in \ref{SupSec:TwoModeLocalGlobal}.

Because the residual operator is proportional to a single lowering operator, the two-mode problem reduces to a scalar full-cycle
constraint. This makes the model the minimal setting in which partial local cancellation with full-cycle flow constraint can be
implemented explicitly.

\subsection{Dressing ansatz and full-cycle flow constraint}

We parameterize the dressing transformation using two time-dependent angles,
\begin{equation} 
    \hS_\mm{dr} (t) = \exp[-i \mu_z (t) \hsigma_z] \exp[-i \mu_x (t) \hsigma_x].
    \label{eq:S_dr_2_mode_ex}
\end{equation} 
The order of the two factors is fixed throughout the calculation and in the numerical solution of the full-cycle constraint.

Each angle is expanded in a truncated Fourier series according to 
\begin{equation}
    \begin{aligned}
        \mu_x (t) &= \frac{\pi}{2 C_x} \sum_l a_{x,l} \left[ 1 - \cos\left( \frac{2\pi l}{t_0} t \right) \right],\\
        \mu_z (t) &= \frac{\pi}{2}\left( 1 + \frac{1}{C_z} \right) \left[  \sum_l
        a_{z,l} \left[ 1 - \cos\left( \frac{2\pi l}{t_0} t \right) \right] \right.\\
        &\phantom{={}}    
        \left.  + \sum_m b_{z,m} \sin\left( \frac{2\pi m}{t_0} t \right) \right],
    \end{aligned}
    \label{eq:DressingAnglesOpt}
\end{equation}
with $C_j=\abs{\sum_l 2 a_{j,l} + \sum_m b_{j,m}}$, $j \in \{x,z\}$. The normalization terms prevent discontinuities from
appearing when the coefficients are varied. This parametrization provides smooth dressing functions and gives $\mu_x(0) =
\mu_x(t_0) = 0$ and $\mu_z(0) = \mu_z(t_0) = \pi/2$. Hence, $\hS_\mm{dr}(0)=\hS_\mm{dr}(t_0)=-i\hsigma_z$. These boundary values
satisfy the conditions in Eq.~\eqref{eq:DressingBoundaryConditions}, while the algebraically determined real control fields
satisfy Eq.~\eqref{eq:ControlBoundaryConditions}.

The Fourier coefficients are free parameters used to enforce the full-cycle constraint on the residual dressed-frame coupling. For
each choice of coefficients, the corresponding fields $g_x (t)$ and $g_z (t)$ are fixed by the local cancellation condition
$\beta_+ (t) = 0$. Thus, after local cancellation, the task is to find dressing coefficients for which the residual
interaction-picture flow has no net effect at the final time. The dressing functions and laboratory-frame controls are therefore
obtained self-consistently. For the solutions used here, the dressing parametrization and the algebraic control relations give
$g_x(0)=g_x(t_0)=g_z(0)=g_z(t_0)=0$, as illustrated in Fig.~\ref{fig:sample_result}c. The modified and bare Hamiltonians
therefore coincide at the endpoints and share the same initial and final instantaneous eigenstates. These physical control-field
boundary conditions are distinct from the boundary conditions on $\hS_\mm{dr}(t)$. The dressing transformation need not reduce to
the identity, provided that its initial and final actions preserve the target subspace and do not alter the desired projected
endpoint map. Together, these properties ensure that the protocol maps instantaneous eigenstates of the original exceptional-point
loop between the same physical endpoints as ideal adiabatic branch exchange.

For the two-mode residual coupling, the interaction-picture Hamiltonian has the form
\begin{equation}
    \hV_\uI (t) = \tilde{\beta}_- (t) \hsigma_-,
    \label{eq:VI2Mode}
\end{equation}
and the requirement that the residual interaction-picture flow have no net effect over the cycle reduces to the scalar constraint
\begin{equation}
    \int_0^{t_0} \di{t} \tilde{\beta}_-(t) = 0.
    \label{eq:Phi_I_int_2mode}
\end{equation}
This equation ensures that the accumulated effect of the residual coupling vanishes over the complete loop, and is equivalent to
imposing Eq.~\eqref{eq:IntPictFlow} for the present two-mode case. It can be solved directly as a nonlinear constraint; in the
numerical results below we impose it through a squared residual, as described in the next subsection. The explicit expression for
$\tilde{\beta}_-(t)$ is given in \ref{SupSec:TwoModeLocalGlobal}.

\subsection{Numerical solution of the full-cycle constraint}

For each loop duration $t_0$ and base point $\varphi$, the Fourier coefficients defining the dressing angles are chosen so that
the residual interaction-picture flow has no net effect at the final time. In general, this gives nonlinear equations for the
dressing parameters. These equations can be solved directly by root finding or with constrained numerical methods when additional
pulse properties are desired.

For the two-mode results reported here, the residual interaction-picture Hamiltonian is proportional to a single lowering
operator, so the full-cycle constraint reduces to the scalar equation in Eq.~\eqref{eq:Phi_I_int_2mode}. We impose this constraint
numerically by minimizing the squared residual
\begin{equation}
    \mathcal{J} = \abs{\int_0^{t_0} \di{t} \tilde{\beta}_-(t)}^2.
    \label{eq:Min2Mode}
\end{equation}
This procedure is used as a robust numerical root-finding method, not as a fidelity optimization. Unless otherwise stated, we use
truncation orders $l=m=3$ in Eq.~\eqref{eq:DressingAnglesOpt} and bound all Fourier coefficients within
$-5 \leq a_{j,l}, b_{j,m} \leq 5$.

For each candidate set of Fourier coefficients defining the dressing angles, the local cancellation condition
$\beta_+(t)=0$ algebraically determines a corresponding pair of real fields $g_x(t)$ and $g_z(t)$. These fields are therefore
functions of the dressing coefficients, not independently prescribed controls. Once a set of coefficients satisfying the
full-cycle constraint is found, the associated $g_x (t)$ and $g_z (t)$ define the laboratory waveform used in the propagation. The
modified dynamics is then obtained by direct numerical propagation of the time-dependent Schrödinger equation generated by
\begin{equation}
    \hH_\mm{mod} (t) = \hH_\mm{sym} (t) + \hW (t),
    \label{eq:Hmod2Mode}
\end{equation}
without further approximation. The final branch-transition probability is not used to define the controls. It is evaluated only
after propagation as a diagnostic of whether the resulting real-control waveform reproduces the target state transfer.

\subsection{Performance and robustness diagnostics}

Performance is quantified using the normalized branch-transition probability
\begin{equation}
    P_{i,j}(t) =
    \frac{\left|\bra{\psi_j}\hPhi (t)\ket{\psi_i}\right|^2}
    {\sum_k \left|\bra{\psi_k}\hPhi (t)\ket{\psi_i}\right|^2}.
    \label{eq:Prob}
\end{equation}
Here, $\hPhi(t)=\hPhi_\mm{ad}(t)$ for the uncontrolled dynamics and $\hPhi(t)=\hPhi_\mm{mod,ad}(t)$ for the constrained-control
dynamics. Both flows are expressed in the same instantaneous-eigenstate frame constructed from the holomorphically continued
eigenvalues and eigenmodes of the bare loop. Thus, $P_{i,j}(t)$ is a normalized branch-transition probability in the
holomorphically continued instantaneous-eigenstate frame. We define the state-transfer error
\begin{equation}
    \varepsilon_{i,j} (t) = 1 - P_{i,j} (t),
    \label{eq:Error}
\end{equation}
so that low final error $\varepsilon_{i,i} (t_0)$ indicates that the controlled dynamics maps the initial instantaneous eigenstate
to the final eigenstate selected by holomorphic continuation of the same branch label. It is a final-state diagnostic and does not
require branch following at intermediate times. For a loop enclosing an exceptional point, this holomorphic continuation defines
the desired exchange of the two instantaneous eigenstates. For the two-dimensional robustness maps in
Fig.~\ref{fig:robustness}e,f, we plot the branch-averaged error
$\bar{\varepsilon}(t_0)=[\varepsilon_{1,1}(t_0)+\varepsilon_{2,2}(t_0)]/2$ to summarize the two transported branches on a single
color scale.

Robustness to errors in the implemented loop is assessed by adding a small static offset to the detuning,
\begin{equation}
    \Delta_0 \to \Delta_0 + \delta \Delta_0 ,
    \label{eq:DetUncert}
\end{equation}
without recomputing the dressing angles or the corresponding coherent-control fields. This tests whether a protocol constructed
for a nominal calibrated loop remains effective when the implemented trajectory is imperfect. The detuning offset $\delta\Delta_0$
is sampled from a Gaussian distribution with zero mean and standard deviation $\sigma = 3\Delta_0/100$, and the state-transfer
error is averaged over a thousand disorder realizations. This perturbation contributes to both diagonal and off-diagonal terms in
the instantaneous-eigenstate frame and is therefore not explicitly cancelled by the coherent-control Hamiltonian.

We also test two additional implementation errors. To model a static calibration error in the inferred exceptional-point position,
we keep the control waveforms fixed and propagate the dynamics with a constant offset in the non-Hermitian scale,
\begin{equation}
    \Gamma_0 \to \Gamma_0 + \delta\Gamma_0.
\end{equation}
Here, $\delta\Gamma_0$ is held fixed during a given propagation and represents a mismatch between the value of $\Gamma_0$ used to
construct the controls and the value realized in the experiment. Because the exceptional-point locations depend on $\Gamma_0$,
this static offset models uncertainty in the calibrated non-Hermitian background. The final state-transfer error is then evaluated using
Eq.~\eqref{eq:Error}. 

To model finite-bandwidth implementation, we pass the two target control waveforms, $\Omega(t)+g_x(t)$ and $\Delta(t)+g_z(t)$,
through identical first-order low-pass filters before using them in the physical Hamiltonian. We denote the filtered waveforms by
$\Omega_\mm{imp} (t)$ and $\Delta_\mm{imp} (t)$. The cutoff frequency $\omega_\uc$ represents the bandwidth of the control
hardware, for frequency components $\omega \ll \omega_\uc$ the implemented waveform follows the target waveform, whereas
components $\omega \gtrsim \omega_\uc$ are attenuated and acquire phase lag. In frequency space,
\begin{equation}
    \begin{aligned}
        \Omega_\mm{imp} (\omega) &= \frac{\Omega (\omega) +g_x (\omega)}{1 + i \frac{\omega}{\omega_\uc}},\\
        \Delta_\mm{imp} (\omega) &= \frac{\Delta (\omega) +g_z (\omega)}{1 + i \frac{\omega}{\omega_\uc}}.
    \end{aligned}
    \label{eq:LowPassFreq}
\end{equation}
Equivalently, in the time domain,
\begin{equation}
    \begin{aligned}
        \dot{\Omega}_\mm{imp} (t) &= \omega_\uc \left[\Omega (t) + g_x (t) -\Omega_\mm{imp} (t)\right],\\
        \dot{\Delta}_\mm{imp} (t) &= \omega_\uc \left[\Delta (t) + g_z (t) -\Delta_\mm{imp} (t)\right].
    \end{aligned}
    \label{eq:LowPassTime}
\end{equation}
The filtered controls are used directly in the Hamiltonian, with $\Omega (t) + g_x(t)$ replaced by $\Omega_\mm{imp} (t)$ and
$\Delta (t) + g_z (t)$ replaced by $\Delta_\mm{imp} (t)$, without recomputing the dressing functions. This tests the sensitivity
of the completed protocol to actuator smoothing and phase lag.

\subsection{Boundary conditions for the dressing transformation}
\label{sec:DressingBoundaryConditions}

We derive sufficient boundary conditions on the dressing transformation that preserve the desired endpoint flow. Let
$\hPhi_\mm{mod,ad} (t_0)$ denote the modified flow expressed in the instantaneous-eigenstate frame of the bare Hamiltonian, and
let $\hPhi_\mm{dr} (t_0)$ denote the corresponding flow in the dressed frame. From the definition of the dressing transformation,
these flows are related by 
\begin{equation} 
    \hPhi_\mm{mod,ad}(t_0) = \hS_\mm{dr}(t_0) \hPhi_\mm{dr}(t_0) \hS_\mm{dr}^{-1}(0). 
    \label{eq:AdDrFlowRelation} 
\end{equation} 
Here, $\hS_\mm{dr}^{-1}(0)$ transforms the initial instantaneous eigenstates into dressed states, $\hPhi_\mm{dr}(t_0)$
generates the dressed-frame flow, and $\hS_\mm{dr}(t_0)$ transforms the final dressed states back into instantaneous eigenstates.
The dressing transformation need not reduce to the identity at either boundary. The relevant requirement is that these initial and
final changes of frame do not alter the desired endpoint flow.

Let $\hP$ project onto the subspace of holomorphically continued states to be transported and let $\hQ=\mathbbm{1}-\hP$. We first
require the two changes of frame in Eq.~\eqref{eq:AdDrFlowRelation} not to mix the target subspace with its complement.
Explicitly,
\begin{equation} 
    \begin{aligned} 
        \hQ \hS_\mm{dr}^{-1}(0) \hP &= \mathbf{0}, \\ 
        \hP \hS_\mm{dr}^{-1}(0) \hQ &= \mathbf{0}, \\ 
        \hQ \hS_\mm{dr}(t_0) \hP &= \mathbf{0}, \\ 
        \hP \hS_\mm{dr}(t_0) \hQ &= \mathbf{0}. 
    \end{aligned} 
    \label{eq:DressingNoSubspaceMixing} 
\end{equation}
The first two lines ensure that the initial transformation into the dressed frame preserves the target subspace, while the last
two ensure that the final transformation back into the instantaneous-eigenstate frame does the same.
Equation~\eqref{eq:DressingNoSubspaceMixing} is equivalently written as
\begin{equation} 
    \begin{aligned} 
        \left[ \hP, \hS_\mm{dr}^{-1}(0) \right] &= \mathbf{0}, \\ 
        \left[ \hP, \hS_\mm{dr}(t_0) \right] &= \mathbf{0}. 
    \end{aligned} 
    \label{eq:DressingPreservesSubspace} 
\end{equation}
Under these conditions, the flow within the target subspace follows from Eq.~\eqref{eq:AdDrFlowRelation} as 
\begin{equation} 
    \hP \hPhi_\mm{mod,ad}(t_0) \hP = \hP \hS_\mm{dr}(t_0) \hP \hPhi_\mm{dr}(t_0) \hP \hS_\mm{dr}^{-1}(0) \hP. 
    \label{eq:ProjectedAdDrFlow} 
\end{equation}
The desired result is that the flow in the instantaneous-eigenstate frame reproduces the corresponding dressed-frame endpoint flow
within the target subspace, 
\begin{equation} 
    \hP \hPhi_\mm{mod,ad}(t_0) \hP = \hP \hPhi_\mm{dr}(t_0) \hP. 
    \label{eq:DesiredProjectedFlow} 
\end{equation}
A sufficient boundary condition is that the final transformation back into the instantaneous-eigenstate frame reverses the initial
transformation into the dressed frame within the target subspace,
\begin{equation} 
    \hP \hS_\mm{dr}(t_0) \hS_\mm{dr}^{-1}(0) \hP = \hP. 
    \label{eq:EqualBoundaryDressing} 
\end{equation} 
Because the two transformations are separated by the dressed-frame flow in Eq.~\eqref{eq:ProjectedAdDrFlow}, we additionally
require the final dressing transformation to commute with the relevant dressed-frame endpoint flow,
\begin{equation} 
    \left[ \hP \hS_\mm{dr}(t_0) \hP,\, \hP \hPhi_\mm{dr}(t_0) \hP \right] = \mathbf{0}. 
    \label{eq:DressingFlowCommutation} 
\end{equation}
Alternatively, under Eq.~\eqref{eq:EqualBoundaryDressing}, the equivalent condition may be imposed on the initial transformation,
\begin{equation} 
    \left[ \hP \hS_\mm{dr}^{-1}(0) \hP,\, \hP \hPhi_\mm{dr}(t_0) \hP \right] = \mathbf{0}. 
    \label{eq:InitialDressingFlowCommutation} 
\end{equation}
Using Eq.~\eqref{eq:DressingFlowCommutation}, the final dressing transformation can be moved through the projected dressed-frame
flow. Equation~\eqref{eq:ProjectedAdDrFlow} then becomes 
\begin{equation} 
    \begin{aligned} 
        \hP \hPhi_\mm{mod,ad}(t_0) \hP &= \hP \hPhi_\mm{dr}(t_0) \hP \hS_\mm{dr}(t_0) \hS_\mm{dr}^{-1}(0) \hP \\ 
        &= \hP \hPhi_\mm{dr}(t_0) \hP, 
    \end{aligned} 
    \label{eq:EqualProjectedFlows} 
\end{equation}
where the second equality follows from Eq.~\eqref{eq:EqualBoundaryDressing}. Thus, the final transformation back into the
instantaneous-eigenstate frame reverses the initial transformation into the dressed frame without altering the desired endpoint
flow. The dressing transformation may be nontrivial at both boundaries, provided that it preserves the target subspace, has the
required combined action at the boundaries and commutes with the relevant dressed-frame endpoint flow.

The boundary conditions in Eq.~\eqref{eq:DressingBoundaryConditions} should be distinguished from those imposed on the physical
control fields. In the two-mode protocols studied here, the constructed fields satisfy 
\begin{equation} 
    \begin{aligned} 
        g_x(0) &= g_x(t_0) = 0, \\ 
        g_z(0) &= g_z(t_0) = 0. 
    \end{aligned} 
    \label{eq:ControlBoundaryConditions} 
\end{equation}
The modified and bare Hamiltonians therefore coincide at the beginning and end of the protocol,
\begin{equation} 
    \begin{aligned} 
        \hH_\mm{mod}(0) &= \hH_\mm{sym}(0), \\ 
        \hH_\mm{mod}(t_0) &= \hH_\mm{sym}(t_0). 
    \end{aligned} 
    \label{eq:HamiltonianBoundaryMatching} 
\end{equation} 
Consequently, the protocol begins with an instantaneous eigenstate of the original exceptional-point loop and ends with the
instantaneous eigenstate selected by holomorphic continuation of that loop. The boundary conditions on the dressing transformation
ensure that the auxiliary changes of frame do not alter this endpoint flow, while the vanishing physical control fields ensure
that the initial and final states are instantaneous eigenstates of the bare Hamiltonian.

\subsection{Pulse-resource metrics}
\label{subsec:pulse_resources}

The RMS-amplitude and bandwidth analyses in Fig.~\ref{fig:robustness} quantify the additional resources required by the modified
real-control loop. Since the relative dissipation $\Gamma_0$ is fixed, the resource cost is computed only from the two tunable
control waveforms. For the uncorrected loop these are $\Omega(t)$ and $\Delta(t)$, while for the constrained-control protocol they
are $\Omega(t)+g_x(t)$ and $\Delta(t)+g_z(t)$.

We define the RMS amplitudes of the uncorrected and constrained-control waveforms as
\begin{equation}
	\begin{aligned}
        A_\mm{rms}^\mm{UC} &= \left[ \frac{1}{t_0} \int_0^{t_0} \di{t}\, \left( \abs{\Omega(t)}^2 + \abs{\Delta(t)}^2 \right) \right]^{1/2},\\
        A_\mm{rms}^\mm{CCP} &= \left[ \frac{1}{t_0} \int_0^{t_0} \di{t}\, \left( \abs{\Omega(t)+g_x(t)}^2 + \abs{\Delta(t)+g_z(t)}^2 \right) \right]^{1/2}.
	\end{aligned}
	\label{eq:Arms}
\end{equation}
The amplitude overhead plotted in Fig.~\ref{fig:robustness}c is
\begin{equation}
    R_\mm{rms} = \frac{A_\mm{rms}^\mm{CCP}}{A_\mm{rms}^\mm{UC}}, 
	\label{eq:Rrms}
\end{equation}
where UC denotes the uncorrected loop and CCP denotes the constrained-control protocol.

For the bandwidth analysis, we first subtract the time average from each control channel. The uncorrected spectral power is
defined from the Fourier transforms of $\Omega(t)$ and $\Delta(t)$,
\begin{equation}
    S_\mm{UC} (\omega) = \abs{\Omega(\omega)}^2 + \abs{\Delta(\omega)}^2,
    \label{eq:SUC}
\end{equation}
whereas the constrained-control spectral power is defined from the Fourier transforms of the modified waveforms,
\begin{equation}
    S_{\rm CCP}(\omega) = \abs{\Omega(\omega) +g_x (\omega)}^2 + \abs{\Delta(\omega) +g_z (\omega)}^2.
    \label{eq:SCCP}
\end{equation}
For each case, the bandwidth $B_{99}$ is the smallest positive frequency satisfying
\begin{equation}
    \frac{ \int_{-B_{99}}^{B_{99}} \di{\omega}\,S(\omega)} { \int_{-\infty}^{\infty} \di{\omega}\,S(\omega)} = 0.99,
    \label{eq:B99}
\end{equation}
with $S(\omega) = S_\mm{UC} (\omega)$ or $S_\mm{CCP}(\omega)$ as appropriate. The bandwidth overhead plotted in
Fig.~\ref{fig:robustness}d is
\begin{equation}
    R_\mm{bw} = \frac{B_{99}^\mm{CCP}}{B_{99}^\mm{UC}}.
    \label{eq:Rbw}
\end{equation}

\npsection{Data availability}

The numerical data used to generate the figures can be reproduced using the code available at~\cite{Chavva2026}.

\npsection{Code availability}

The code used to generate the control waveforms, perform the time-dependent propagation and produce the figures is available at~\cite{Chavva2026}.

\npsection{Author contributions}

V.C. and N.O. performed the analytical calculations and numerical simulations. H.R. conceived and supervised the project.  All
authors discussed the results and contributed to writing the manuscript.

\def\bibsection{\npsection{References}}
\bibliography{RealFields.bib}

\clearpage
\onecolumngrid
\allowdisplaybreaks

\npsection{Supplementary Information}

\setcounter{suppnote}{0}
\setcounter{equation}{0}
\renewcommand{\theequation}{S\arabic{equation}}
\setcounter{figure}{0}
\renewcommand{\thefigure}{\arabic{figure}}
\setcounter{table}{0}
\renewcommand{\thetable}{S\arabic{table}}
\addto\captionsenglish{\renewcommand{\figurename}{Supplementary Figure}}

\suppnote{Holomorphic continuation and construction of the instantaneous-eigenstate frame}
\label{SupSec:ChangeOfFrame}

This Supplementary Note defines the holomorphically continued instantaneous-eigenstate frame used throughout the manuscript and
explains why following a fixed branch in this frame implements the endpoint state transfer selected by the exceptional-point
braid.

\subsection*{Instantaneous eigenstates and biorthogonal basis}

We consider a non-Hermitian Hamiltonian $\hH[\pv (t)]$, where $\pv (t)$ denotes a set of externally controlled parameters. At each
time $t$, the right and left instantaneous eigenstates are defined as 
\begin{equation}
    \begin{aligned}
        \hH[\pv (t)] \ket{\psi_j (t)} &= \lambda_j (t) \ket{\psi_j (t)},\\
        \bra{\psi_j (t)} \hH[\pv (t)] &= \lambda_j (t) \bra{\psi_j (t)},
    \end{aligned}
    \label{eq:LeftRightEigVecs}
\end{equation}
with complex eigenvalues $\lambda_j (t)$. The right and left eigenstates are chosen to satisfy the biorthogonal normalization 
condition
\begin{equation}
    \braket{\psi_j (t)}{\psi_i (t)} = \delta_{i,j},
    \label{eq:BiorthogonalNorm}
\end{equation}
where $\delta_{i,j}$ is the Kronecker delta function.

The instantaneous‑eigenstate frame is defined by the generally non‑unitary change‑of‑frame operator
\begin{equation}
    \hS_\mm{ad} (t) = \sum_j \ketbra{\psi_j (t)}{\psi_j}. 
    \label{eq:Sad}
\end{equation}
In the instantaneous-eigenstate frame, the eigenvectors $\ket{\psi_j}$ are static.

\subsection*{Holomorphic continuation around an exceptional point}

When the parameter loop $\pv (t)$ winds around an exceptional point, the eigenvalues $\lambda_j (t)$ and eigenstates $\ket{\psi_j
(t)}$ are multi‑valued functions of the parameters. To define the instantaneous eigenstates consistently along the loop, we follow
each eigenvalue and eigenstate by holomorphic continuation in parameter space~\cite{chavva2025_2}. 

Concretely, the eigenstates at time $t$ are not defined independently at each point in parameter space, but are obtained by
smoothly continuing the initial eigenstates along the chosen loop. This procedure fixes the branch of the Riemann surface
associated with each eigenvalue and eigenmode.

As a result, after one full loop around the exceptional point, the holomorphically continued eigenmodes need not coincide with the
eigenmodes obtained by diagonalizing $\hH[\pv (0)]$  anew. Instead, they are related by the characteristic exceptional‑point
braid, such as a mode exchange in the two‑mode case.

\subsection*{Adiabatic following and topological transport}

With this construction, the notion of ``remaining on a given adiabatic state'' refers to remaining on a fixed holomorphically continued
eigenmode throughout the protocol. Because the holomorphic continuation encodes the topology of the exceptional point, adiabatic
following in this frame directly enacts the associated topological braid.

In particular, if the system remains on the holomorphically continued eigenmode $\ket{\psi_j(t)}$ for all $t \in [0,t_0]$, then
at the end of the loop the physical state corresponds to the eigenmode obtained by transporting $\ket{\psi_j(0)}$ around the
exceptional point along the chosen branch of the Riemann surface. In the two‑mode setting, this appears as a mode exchange in the
physical eigenbasis.

In the ideal adiabatic limit, following a fixed holomorphically continued branch in the instantaneous-eigenstate frame implements
the endpoint state transfer associated with the exceptional-point braid.

\subsection*{Adiabatic frame Hamiltonian}

In the instantaneous‑eigenstate frame, the Hamiltonian takes the form
\begin{equation}
    \begin{aligned}
        \hH_\mm{ad} (t) &= \hS_\mm{ad}^{-1} (t) \hH[\pv (t)] \hS_\mm{ad} (t) - i \hS_\mm{ad}^{-1} (t) \rd_t \hS_\mm{ad} (t) \\
        &= \hH_{\mm{ad},0} (t) + \hV_\mm{nad} (t),
    \end{aligned}
    \label{eq:HadSupMat}
\end{equation}
where
\begin{equation}
    \hH_0 (t) = \sum_j (\lambda_j (t) -i \braket{\psi_j (t)}{\rd_t \psi_j (t)})   \ketbra{\psi_j}{\psi_j},
    \label{eq:H0Ad}
\end{equation}
is a diagonal operator and 
\begin{equation}
    \hV_\mm{nad} (t) =  -i \sum_{i\neq j} \braket{\psi_j (t)}{\rd_t \psi_i (t)} \ketbra{\psi_j}{\psi_i},
    \label{eq:VNotAd}
\end{equation}
describes the coupling between the eigenmodes generated by the time dependence of the eigenbasis.

In Hermitian systems, sufficiently slow driving suppresses $ \hV_\mm{nad} (t)$. In non‑Hermitian systems, however, these couplings
generally do not vanish even in the slow‑driving limit, which is the origin of the breakdown of the adiabatic theorem near
exceptional points. The control strategies developed in the main text are designed to address this breakdown while preserving the
holomorphic‑continuation structure described above.

\subsection*{Relation to state-transfer error and branch tracking}

The main text defines the state-transfer error as
\begin{equation}
    \varepsilon_{i,j}(t)=1-P_{i,j}(t),
    \label{eq:SupError}
\end{equation}
where $P_{i,j}(t)$ is computed in the holomorphically continued instantaneous-eigenstate frame described here.  Throughout the
manuscript, $P_{i,j}(t)$ denotes a branch-transition probability with respect to the holomorphically continued instantaneous
eigenmodes constructed here. A high final value of $P_{i,i}(t_0)$, or equivalently a low final error $\varepsilon_{i,i}(t_0)$,
indicates that the controlled dynamics maps the initial instantaneous eigenstate to the final eigenstate selected by holomorphic
continuation of the bare loop. This is a final-state diagnostic and does not imply that the physical state follows the
corresponding branch at intermediate times.

\subsection*{Two-mode construction}

For the two-mode case considered in the main text, the change of frame operator can be expressed as 
\begin{equation}
    \hS_\mm{sym,ad}(t) = \exp \left[-\frac{i}{2} \theta (t) \hat{\sigma}_y\right],
    \label{eq:S_ad_holo}
\end{equation}
where
\begin{equation}
        \theta (t) = -2\arctan\left(\frac{\Omega (t)}{\Delta (t) + i \frac{\Gamma_0}{2}+\sqrt{\left(\Delta (t) + i \frac{\Gamma_0}{2} \right)^{2} +
        \Omega^{2} (t)}}\right) + \chi (t).
    \label{eq:cont_theta_static}
\end{equation}
is interpreted as a pseudo-rotation angle. The function $\chi (t)$ describes the holomorphic continuation of the $\arctan$
function leading to holomorphic eigenvalues. For the bare control loop considered in the main text [see Eq.~\eqref{eq:2ModeContour}], this function takes the simple form~\cite{chavva2025_2}
\begin{equation}
    \chi(t)= \pi \Theta \left[ \frac{1-d}{2}+ \left(s(t) - \frac{1}{2} +\frac{\varphi}{2\pi}\right) \right],
    \label{eq:ChiEpsilon}
\end{equation}
where $d$ is the orientation of the loop ($\pm 1$ for clockwise/anticlockwise), $\varphi$ is the base point, and $\Theta$ is the
Heaviside function with convention $\Theta(0)=1$. 

\suppnote{Two-mode model:~native control manifold and transformation of the control Hamiltonian}
\label{SupSec:TwoMode}

This Supplementary Note derives the algebraic constraints imposed by restricting the laboratory-frame controls to the native
control manifold of the original Hamiltonian. Although the corresponding laboratory control fields are real-valued, their matrix
elements in the instantaneous-eigenstate frame are generally complex.

The constrained-control protocol is formulated within the native laboratory-frame control manifold. For the two-mode Hamiltonian
used in the main text, the allowed control Hamiltonian is
\begin{equation}
    \hW_\mm{sym} (t) = g_z (t) \hat{\sigma}_z + g_x (t) \hat{\sigma}_x,
    \label{eq:WsymSupMat}
\end{equation}
with $g_z (t)$, $g_x (t) \in \mathbbm{R}$, as described in the Methods Section and in Eq.~\eqref{eq:Wsym} of the main text. This
restriction allows the protocol to reshape only the native coherent detuning and coupling terms. It does not introduce a
$\hsigma_y$ component, auxiliary dissipation or any non-native coupling.

After transformation to the instantaneous-eigenstate frame,
\begin{equation}
    \hW_\mm{ad} (t) = \hS^{-1}_\mm{ad} (t) \hW_\mm{sym} (t) \hS_\mm{ad} (t),
    \label{eq:WAdSupMat}
\end{equation}
the control Hamiltonian generally has complex matrix elements because $\hS_\mm{ad} (t)$ is non-unitary. The restriction to the
native control manifold is therefore enforced not by requiring $\hW_\mm{ad}(t)$ to have real matrix elements, but by requiring
their real and imaginary parts to satisfy algebraic relations that reconstruct the real-valued native fields $g_x(t)$ and
$g_z(t)$.

\subsection*{Constrained instantaneous-eigenstate frame control coefficients}

Given the forms of the constrained Hamiltonian [see Eq.~\eqref{eq:WsymSupMat}] and change of frame operator [see
Eq.~\eqref{eq:S_ad_holo}], we can expand the transformed control $\hW_\mm{ad} (t)$ in the Pauli basis as
\begin{equation}
    \hW_\mm{ad} (t) =  w_x (t) \hat{\sigma}_x + w_z (t) \hat{\sigma}_z,
    \label{eq:WAdSupMatPauli}
\end{equation}
where the coefficients $w_j(t)$, $j\in\{x,z\}$, are generally complex. Their real and imaginary parts obey algebraic relations
imposed by the two-dimensional native laboratory control manifold spanned by $\hsigma_x$ and $\hsigma_z$. These relations can be
found by transforming Eq.~\eqref{eq:WAdSupMatPauli} back to the laboratory frame using Eq.~\eqref{eq:S_ad_holo}. By imposing that
the matrix elements of the transformed control Hamiltonian $\hW_\mm{sym} (t) = \hS_\mm{ad} (t) \hW_\mm{ad} (t) \hS_\mm{ad}^{-1}
(t)$ must be real, we find 
\begin{equation}
    \begin{aligned}
        \mm{Im}[w_x (t)] &= -\mm{Re}[w_z (t)] \tanh\{\mm{Im}[\theta (t)]\}, \\
        \mm{Im}[w_z (t)] &= \mm{Re}[w_x (t)] \tanh\{\mm{Im}[\theta (t)]\},
    \end{aligned}
    \label{eq:AlgRelWAd}
\end{equation}
where $\mm{Re}[\cdot]$ and $\mm{Im}[\cdot]$ denote the real and imaginary parts of a complex quantity.

The fields $g_x(t)$ and $g_z(t)$ can then be reconstructed as
\begin{equation}
    \begin{aligned}
        g_x (t) = \sech\left\{ \mm{Im}\left[ \theta (t) \right] \right\} \left( \cos\left\{ \mm{Re}\left[\theta (t)\right]\right\} 
            \mm{Re}[w_x (t)] + \sin\left\{ \mm{Re}\left[\theta (t)\right] \right\} \mm{Re}[w_z (t)] \right),\\
        g_z (t) = \sech\left\{ \mm{Im}\left[ \theta (t) \right] \right\} \left( \cos\left\{ \mm{Re}\left[\theta (t)\right] \right\} 
            \mm{Re}[w_z (t)] - \sin\left\{ \mm{Re}\left[\theta (t)\right] \right\} \mm{Re}[w_x (t)] \right),
    \end{aligned}
    \label{eq:ParamCtrlFields}
\end{equation}
where $\mm{Re}[w_x(t)]$ and $\mm{Re}[w_z(t)]$ encode the two independent control degrees of freedom available within the native
control manifold for local cancellation of dressed-frame coupling channels.

Equations~\eqref{eq:WAdSupMatPauli} and \eqref{eq:AlgRelWAd} allow us to conveniently work in the adiabatic frame and then
reconstruct the laboratory fields through the relations given in Eq.~\eqref{eq:ParamCtrlFields}.

\subsection*{Role in the coherent-control procedure}

The constrained-control protocol uses the two available real fields $g_x(t)$ and $g_z(t)$ to cancel the dressed-frame
coupling channel that is accessible under the real-control constraint. The remaining dressed-frame coupling channel is not forced
to vanish instantaneously. Instead, the dressing angles are chosen so that the residual interaction-picture flow has no net effect
after one full cycle.

In this sense, the real-control constraint plays two roles. First, it determines which combinations of dressed-frame matrix
elements can be cancelled locally. Second, for any candidate dressing, it algebraically determines the corresponding
laboratory-frame waveforms $g_x(t)$ and $g_z(t)$. The final waveforms are obtained once the dressing coefficients satisfy the
full-cycle flow constraint.

\suppnote{Conventional dressed-state shortcuts to adiabaticity and the overconstraint from native real controls}
\label{SupSec:DressedSTA}

This Supplementary Note shows why complete instantaneous cancellation is not the natural constrained objective within the native
control manifold considered in the main text. In the conventional dressed-state shortcut-to-adiabaticity (STA) construction, all
off-diagonal dressed-frame couplings are required to vanish locally. The original construction also imposes the no-dressing
requirement $\hS_\mm{dr}(0)=\hS_\mm{dr}(t_0)=\mathbbm{1}$, so that the dressed instantaneous eigenstates coincide with the
instantaneous eigenstates at the beginning and end of the protocol. Within the two-dimensional native control manifold, however,
only one dressed-frame transition channel can be cancelled independently. Enforcing complete local cancellation therefore
transfers the remaining conditions to the dressing parameters, producing coupled nonlinear differential equations that become
singular or ill-conditioned near the exceptional point.

The dressed-state approach requires one to choose a control Hamiltonian $\hW_\mm{sym}(t)$ and a dressing transformation
$\hS_\mm{dr}(t)$ so that all off-diagonal couplings between dressed states vanish at every instant. For the two-mode non-Hermitian
Hamiltonian considered here, this instantaneous-cancellation condition imposes more real constraints than can be satisfied
independently by the available controls. The remaining conditions can, in principle, be absorbed into the dressing transformation.
However, cancelling all dressed-frame transition channels locally while enforcing the no-dressing requirement at the protocol
boundaries and restricting $\hW_\mm{sym}(t)$ to its native form with $g_x(t),g_z(t)\in\mathbbm{R}$ leads to coupled nonlinear
ordinary differential equations for the dressing parameters. These equations become singular or ill-conditioned near the
exceptional point.

We illustrate this obstruction using the dressing transformation employed in the main text [see Eq.~\eqref{eq:S_dr_2_mode_ex}],
\begin{equation}
    \hS_\mm{dr} (t) = \exp[-i \mu_z (t) \hsigma_z] \exp[-i \mu_x (t) \hsigma_x].
    \label{eq:S_dr_2mode_sup}
\end{equation}
Starting from the modified laboratory-frame non-Hermitian Hamiltonian $\hH_\mm{mod}(t) = \hH_\mm{sym} (t) + \hW_\mm{sym} (t)$, with
$\hH_\mm{sym} (t)$ given in Eq.~\eqref{eq:H_sym} of the main text and $\hW_\mm{sym} (t)$ in Eq.~\eqref{eq:WsymSupMat}, we
successively transform the modified Hamiltonian to the adiabatic and dressed frames using Eqs.~\eqref{eq:S_ad_holo} and
\eqref{eq:S_dr_2mode_sup}, respectively. The resulting dressed-frame Hamiltonian is 
\begin{equation}
    \hH_\mm{dr} (t) = Z (t) \hat{\sigma}_z + X(t) \hat{\sigma}_x + Y(t) \hat{\sigma}_y,
    \label{eq:H_dr_Pauli_decomp}
\end{equation}
with 
\begin{equation}
    \begin{aligned}
        X (t) &= \frac{1}{2} \left(2 \cos \mu_z(t) \left\{g_x(t)-i g_z(t) \tanh\{\mm{Im}[\theta(t)]\}\right\}\right)- \sin \mu_z(t) \left[\dot\theta(t)+\dot\mu_x(t) \right],\\
        Y (t) &= \frac{1}{2} \left\{\vphantom{\dot\theta} \sin \mu_x \left\{ 2 \lambda(t)-\dot\mu_z(t) + 2i g_x(t) \tanh\{\mm{Im}[\theta(t)]\}\right\}-\cos \mu_x (t) \cos \mu_z (t) \dot\theta(t)\vphantom{\left\{-2 g_x(t) \cos \mu_x(t) \sin \mu_z(t) + 2 g_z(t) \left\{\sin \mu_x(t)+ i \cos \mu_x(t) \sin \mu_z (t) \tanh\{\mm{Im}[\theta (t)]\}\right\}\right.} \right.\\
        &\phantom{={}} \left. -2 g_x(t) \cos \mu_x(t) \sin \mu_z(t) + 2 g_z(t) \left\{\sin \mu_x(t)+ i \cos \mu_x(t) \sin \mu_z (t) \tanh\{\mm{Im}[\theta (t)]\}\right\} \vphantom{\dot\theta}\right\},\\
        Z (t) &= \frac{1}{2} \left\{\vphantom{\dot\theta}2 g_x(t) \sin \mu_x(t) \sin \mu_z(t) + 2 g_z(t) \left\{\cos \mu_x(t)- i \sin \mu_x(t) \sin \mu_z (t) \tanh\{\mm{Im}[\theta (t)]\}\right\} \right.\\
        &\phantom{={}} \left. + \cos \mu_x \left\{ 2 \lambda(t)-\dot\mu_z(t) + 2i g_x(t) \tanh\{\mm{Im}[\theta(t)]\}\right\}+\cos \mu_z (t) \sin \mu_z (t) \dot\theta(t)\vphantom{\left\{2 g_x(t) \sin \mu_x(t) \sin \mu_z(t) + 2 g_z(t) \left\{\cos \mu_x(t)- i \sin \mu_x(t) \sin \mu_z (t) \tanh\{\mm{Im}[\theta (t)]\}\right\}\right.}\right\}.
    \end{aligned}
    \label{eq:Coeffs2ModDrH}
\end{equation}
The goal of the conventional dressed-state construction is to find $g_x(t)$, $g_z(t)$, $\mu_x(t)$ and $\mu_z(t)\in\mathbbm{R}$ such
that $X(t)=Y(t)=0$ at every instant. In this example, we first cancel $\mm{Re}[X(t)]$ and $\mm{Re}[Y(t)]$ using the real control
fields $g_x(t)$ and $g_z(t)$. This gives
\begin{equation}
    \begin{aligned}
        g_x (t) &= \frac{1}{2} \sec \mu_z (t) \left[\sin \mu_z(t) \mm{Re}[\dot\theta(t)]+\dot\mu_x(t) \right],\\
        g_z (t) &= \frac{1}{2} \left\{-2\lambda_r (t) + \cot \mu_x(t) \left[\mm{Re}[\dot\theta(t)]+\dot\mu_x(t)\sin \mu_z(t) + \dot\mu_z(t) \right] \right\}.
    \end{aligned}
    \label{eq:CtrlFieldsDSAEx}
\end{equation}
Substituting Eq.~\eqref{eq:CtrlFieldsDSAEx} into Eq.~\eqref{eq:Coeffs2ModDrH} and imposing $\mm{Im}[X(t)]=\mm{Im}[Y(t)]=0$ gives
the following coupled nonlinear ordinary differential equations for $\mu_x(t)$ and $\mu_z(t)$:
\begin{equation}
    \begin{aligned}
        \dot{\mu}_x (t) &= \cot \mm{Im}[\theta(t)] \left [-2 \mm{Im}[\lambda(t)] \cos \mu_z(t) + \mm{Im}[\dot\theta(t)] \cot \mu_x(t)  \right] - \mm{Re}[\dot\theta(t)] \sin \mu_z(t),\\
        \dot{\mu}_z (t) &= \cot \mm{Im}[\theta(t)] \left \{2 \mm{Im}[\lambda(t)] \cot \mu_x(t) \sin \mu_z(t) - \mm{Im}[\dot\theta(t)] \tan \mu_z(t) \left[\csc \mu_x(t)\right]^2  \right\} \\
        &\phantom{={}} + 2 \mm{Re}[\lambda(t)]
    \end{aligned}
    \label{eq:DiffEqDrAnglesDSA}
\end{equation}
The original dressed-state STA construction imposes the no-dressing requirement
\begin{equation}
    \hS_\mm{dr}(0) = \hS_\mm{dr} (t_0) = \mathbbm{1}.
    \label{eq:SupOriginalDressingBoundaryConditions}
\end{equation}
For the parametrization in Eq.~\eqref{eq:S_dr_2mode_sup}, these conditions are implemented by
\begin{equation}
    \begin{aligned}
        \mu_x(0) &= \mu_x(t_0) = 0, \\
        \mu_z(0) &= \mu_z(t_0) = 0.
    \end{aligned}
    \label{eq:SupOriginalAngleBoundaryConditions}
\end{equation}
These no-dressing conditions ensure that the instantaneous eigenstates are not dressed at the beginning or end of the
protocol:~the dressed instantaneous eigenstates therefore coincide with the instantaneous eigenstates at both boundaries. The
conventional dressed-state construction examined in this Supplementary Note imposes this requirement. The present protocol uses
more general boundary conditions and does not require $\hS_\mm{dr}(0)$ or $\hS_\mm{dr}(t_0)$ to reduce to the identity. Instead,
the action of the dressing transformation at the boundaries must preserve the target subspace and leave the desired projected
endpoint flow unchanged.

Solutions of the resulting equations may exist for specially designed trajectories, but their singular or ill-conditioned form
reflects the control-manifold mismatch discussed in the main text. The two independent control degrees of freedom within the
native manifold can cancel only one of the two dressed-frame transition channels locally. The conventional construction
nevertheless requires the second channel to vanish instantaneously and therefore transfers this additional cancellation
requirement to the dressing angles. Combined with the no-dressing requirement in
Eq.~\eqref{eq:SupOriginalDressingBoundaryConditions}, this produces coupled nonlinear differential equations that become singular
or ill-conditioned near the exceptional point.

This calculation therefore shows why complete local cancellation is not the appropriate constrained objective within the native
control manifold. The singular or ill-conditioned equations are not merely consequences of an inconvenient parametrization or the
absence of a closed-form solution. They arise because complete instantaneous cancellation imposes more independent requirements
than can be satisfied directly within the native control manifold. The present protocol therefore changes the STA condition
itself. One dressed-frame channel is cancelled locally using the available native controls, while the dressing parameters are
chosen so that the transitions accumulated from the remaining channel vanish over the full cycle.

\suppnote{Two-mode implementation of partial local cancellation and full-cycle flow constraint}
\label{SupSec:TwoModeLocalGlobal}

This Supplementary Note gives the explicit two-mode implementation of the constrained-control procedure, including the
dressed-frame decomposition, the locally cancelled coupling channel and the residual coupling channel whose interaction-picture
flow is constrained to have no net effect over the full loop.

Using the dressing transformation defined in Eq.~\eqref{eq:S_dr_2mode_sup} and the control Hamiltonian $\hW_\mm{sym} (t)$
introduced in Eq.~\eqref{eq:WsymSupMat}, with the native control fields $g_x (t)$ and $g_z (t)$ parametrized according to
Eq.~\eqref{eq:ParamCtrlFields}, we find that the dressed-frame non-Hermitian Hamiltonian is given by
\begin{equation}
    \hH_\mm{dr} (t) = Z(t) \hat{\sigma}_z + \beta_+ (t) \hat{\sigma}_+ + \beta_- (t) \hat{\sigma}_-
    \label{eq:HModDr2Mode}
\end{equation}
where $Z(t)$ is given in Eq.~\eqref{eq:Coeffs2ModDrH}, and with
\begin{equation}
    \begin{aligned}
        \beta_+ (t) & = \left(\vphantom{\left[\sin \mu_z(t) - \cos \mu_x(t) \cos \mu_z(t) \dot\theta(t) \right]} 2 g_x(t) \left[1-\cos\mu_x(t) +e^{i 2 \mu_z(t)} \left[1+\cos\mu_x(t)\right] +2e^{i  \mu_z(t)} \sin\mu_x(t)\tanh\{\mm{Im}[\theta(t)]\}\right] \right.\\
        &\phantom{={}} \left.+2 e^{i \mu_z(t)} \left\{\left[\sin \mu_z(t) - \cos \mu_x(t) \cos \mu_z(t) \dot\theta(t) \right] \right.\right.\\
        &\phantom{={}} \left.\left.+2 g_z(t) \left\{\sin\mu_x(t)+\left[\cos \mu_z(t)+i \cos \mu_x(t) \sin \mu_z(t)\right]\tanh\{\mm{Im}[\theta(t)]\}\right\}-i \dot\mu_x(t)\right.\right.\\
        &\phantom{={}} \left.\left.+\sin \mu_x(t) \left[2 \lambda (t) - \dot\mu_z(t)\right]\vphantom{\left\{\left[\sin \mu_z(t) - \cos \mu_x(t) \cos \mu_z(t) \dot\theta(t) \right]\right.}\right\}\vphantom{\left( 2 g_x(t) \left[1-\cos\mu_x(t) +e^{i 2 \mu_z(t)} \left[1+\cos\mu_x(t)\right] +2e^{i  \mu_z(t)} \sin\mu_x(t)\tanh\{\mm{Im}[\theta(t)]\}\right]\right.}\right) \frac{1}{4} e^{-i \mu_z(t)},\\
        \beta_- (t) & = \left(2 \left[1+\cos \mu_x(t)\right] g_x(t) -4 i g_z(t) \tanh\{\mm{Im}[\theta(t)]\} \left[\cos \frac{\mu_x(t)}{2}\right]^2 \right.\\
        &\phantom{={}} \left.-i\dot\theta(t)\left[1+\cos\mu_x(t)\right]+2 i e^{i \mu_z(t)}+2 e^{2 i \mu_z(t)}\left[\sin \frac{\mu_x(t)}{2}\right]^2\right.\\
        &\phantom{={}} \left.\times \left\{2 g_x(t)-2ig_z(t)\tanh\{\mm{Im}[\theta(t)]\}-\mm{Im}[\dot\theta(t)]+i \mm{Re}[\dot\theta(t)]\right\} \vphantom{\left(2 \left[1+\cos \mu_x(t)\right] g_x(t) -4 i g_z(t) \tanh\{\mm{Im}[\theta(t)]\} \left[\cos \frac{\mu_x(t)}{2}\right]^2\right.}\right)\frac{1}{4} e^{-i \mu_z(t)}.
    \end{aligned}
    \label{eq:CoeffsHModDr2Mode}
\end{equation}
Here, we choose to cancel the coefficient $\beta_+ (t)$ at all times by choosing [see Eq.~\eqref{eq:ParamCtrlFields}]
\begin{equation}
    \begin{aligned}
        \mm{Re}[w_x (t)] &=\frac{\mm{N}_x(t)}{\mm{D}_x(t)},\\
        \mm{Re}[w_z (t)] &=\frac{\mm{N}_z(t)}{\mm{D}_z(t)},
    \end{aligned}
    \label{eq:CancellationBetaPlus}
\end{equation}
such that
\begin{equation}
    \begin{aligned}
    \mm{N}_x(t) &= \left[\tanh\{\mm{Im}[\theta(t)]\}\cos\mu_z(t)+\sin\mu_x(t)\right]
    \left[-\mm{Im}[\dot\theta(t)]\cos\mu_x(t)\cos\mu_z(t)-\mm{Re}[\dot\theta(t)]\sin\mu_z(t) \right.\\
    &\phantom{={}} \left.+2\mm{Im}[\lambda(t)]\sin\mu_x(t)-\dot\mu_x(t)\vphantom{\left[-\mm{Im}[\dot\theta(t)]\cos\mu_x(t)\cos\mu_z(t)-\mm{Re}[\dot\theta(t)]\sin\mu_z(t)\right.}\right]
    +\tanh\{\mm{Im}[\theta(t)]\}\cos\mu_x(t)\sin\mu_z(t)
    \left(-\mm{Im}[\dot\theta(t)]\sin\mu_z(t) \right.\\
    &\phantom{={}} \left.+\mm{Re}[\dot\theta(t)]\cos\mu_x(t)\cos\mu_z(t)+\sin\mu_x(t)\left\{\dot\mu_z(t)-2\mm{Re}[\lambda(t)]\right\}\vphantom{\left(-\mm{Im}[\dot\theta(t)]\sin\mu_z(t)\right.}\right),\\
    \mm{D}_x(t)&= 2\left\{\tanh\{\mm{Im}[\theta(t)]\}\cos^2\mu_x(t)\sin^2\mu_z(t) \right.\\
    &\phantom{={}} \left.+\left[\tanh\{\mm{Im}[\theta(t)]\}\cos\mu_z(t)+\sin\mu_x(t)\right]\left[\tanh\{\mm{Im}[\theta(t)]\}\sin\mu_x(t)+\cos\mu_z(t)\right]\vphantom{\left\{\tanh\{\mm{Im}[\theta(t)]\}\cos^2\mu_x(t)\sin^2\mu_z(t)\right.}\right\},\\
    \end{aligned}
\end{equation}
and
\begin{equation}
\allowdisplaybreaks
    \begin{aligned}
    \mm{N}_z(t)&= \coth\{\mm{Im}[\theta(t)]\}\left\{-\mm{Im}[\dot\theta(t)]\coth\{\mm{Im}[\theta(t)]\}\sec^2\mu_x(t)\cot\mu_z(t)\right.\\
    &\phantom{={}} \left.-\mm{Im}[\dot\theta(t)]\tan\mu_x(t)\sec\mu_x(t)\csc\mu_z(t)+\mm{Im}[\dot\theta(t)]\coth\{\mm{Im}[\theta(t)]\}\cot\mu_z(t\right.\\
    &\phantom{={}} \left.+\coth\{\mm{Im}[\theta(t)]\}\mm{Re}[\dot\theta(t)]\sec\mu_x(t)\cot^2\mu_z(t)+\coth\{\mm{Im}[\theta(t)]\}\mm{Re}[\dot\theta(t)]\sec\mu_x(t)\right.\\
    &\phantom{={}} \left.-2\mm{Im}[\lambda(t)]\coth\{\mm{Im}[\theta(t)]\}\tan\mu_x(t)\csc\mu_z(t)+\coth\{\mm{Im}[\theta(t)]\}\dot\mu_x(t)\sec\mu_x(t)\csc\mu_z(t)\right.\\
    &\phantom{={}} \left.-2\mm{Re}[\lambda(t)]\tan\mu_x(t)\csc\mu_z(t)\left[\coth\{\mm{Im}[\theta(t)]\}\sec\mu_x(t)\cot\mu_z(t)+\tan\mu_x(t)\csc\mu_z(t)\right] \right.\\
    &\phantom{={}} \left.+\tan\mu_x(t)\dot\mu_z(t)\csc\mu_z(t)\left[\coth\{\mm{Im}[\theta(t)]\}\sec\mu_x(t)\cot\mu_z(t)+\tan\mu_x(t)\csc\mu_z(t)\right] \right.\\
    &\phantom{={}} \left.+\mm{Re}[\dot\theta(t)]\tan\mu_x(t)\cot\mu_z(t)\csc\mu_z(t)\vphantom{\left\{-\mm{Im}[\dot\theta(t)]\coth\{\mm{Im}[\theta(t)]\}\sec^2\mu_x(t)\cot\mu_z(t)\right.}\right\},\\
    \mm{D}_z(t)&= 2\left\{\coth\{\mm{Im}[\theta(t)]\}\left[\sec^2\mu_x(t)\cot^2\mu_z(t)+1\right]+\coth\{\mm{Im}[\theta(t)]\}\tan^2\mu_x(t)\csc^2\mu_z(t) \right.\\
    &\phantom{={}} \left.+\left[\coth^2\{\mm{Im}[\theta(t)]\}+1\right]\tan\mu_x(t)\sec\mu_x(t)\cot\mu_z(t)\csc\mu_z(t)\vphantom{\left\{\coth\{\mm{Im}[\theta(t)]\}\left[\sec^2\mu_x(t)\cot^2\mu_z(t)+1\right]+\coth\{\mm{Im}[\theta(t)]\}\tan^2\mu_x(t)\csc^2\mu_z(t)\right.}\right\}.
    \end{aligned}
\end{equation}
The residual channel
\begin{equation}
    \hV_{\bar{l}} (t) = \beta_- (t) \hat{\sigma}_-,
    \label{eq:Res2ModeEx}
\end{equation}
is not cancelled locally. Instead, the dressing angles are chosen so that the interaction-picture flow generated by this residual
channel is trivial after one full cycle.  This is done by choosing appropriate dressing angles $\mu_x (t)$ and $\mu_z (t)$  for
the dressing operator $\hS_\mm{dr} (t)$ [see Eq.~\eqref{eq:S_dr_2mode_sup}].

\subsection*{Dressing-angle ansatz}

As discussed in the Methods section of the main text, the dressing angles $\mu_x (t)$ and $\mu_z (t)$ are parametrized
using truncated Fourier series
\begin{equation}
    \begin{aligned}
        \mu_x (t) &= \frac{\pi}{2 C_x} \sum_l a_{x,l} \left[ 1 - \cos\left( \frac{2\pi l}{t_0} t \right) \right],\\
        \mu_z (t) &= \frac{\pi}{2}\left( 1 + \frac{1}{C_z} \right) \left[ \sum_l a_{z,l} \left[ 1 - \cos\left( \frac{2\pi l}{t_0} t \right) \right]
            + \sum_m b_{z,m} \sin\left( \frac{2\pi m}{t_0} t \right) \right],
    \end{aligned}
    \label{eq:DressingAnglesOptSup}
\end{equation}
with $C_j=\abs{\sum_l 2 a_{j,l} + \sum_m b_{j,m}}$, $j \in \{x,z\}$. The normalization coefficients prevent discontinuities as the
Fourier coefficients are varied and avoid singular values in the expression for the residual coupling. The Fourier parametrization
is chosen such that
\begin{equation}
    \begin{aligned}
        \mu_x(0) &= \mu_x(t_0) = 0, \\
        \mu_z(0) &= \mu_z(t_0) = \frac{\pi}{2}.
    \end{aligned}
    \label{eq:SupDressingAngleBoundaryValues}
\end{equation}
The dressing transformation therefore has the same, generally nontrivial, value at the two boundaries,
\begin{equation}
    \begin{aligned}
        \hS_\mm{dr}(0) &= \hS_\mm{dr}(t_0) \\
        &= \exp\left[ -i\frac{\pi}{2}\hsigma_z \right] \\
        &= -i\hsigma_z.
    \end{aligned}
    \label{eq:SupDressingBoundaryValues}
\end{equation}
Thus, the present construction does not satisfy the no-dressing requirement $\hS_\mm{dr}(0)=\hS_\mm{dr}(t_0)=\mathbbm{1}$ imposed
in the conventional dressed-state construction discussed in \ref{SupSec:DressedSTA}. Instead, the dressing transformation has the
same nontrivial value at both boundaries. The modified flow expressed in the instantaneous-eigenstate frame is
\begin{equation}
    \hPhi_\mm{mod,ad}(t_0) = \hS_\mm{dr}(t_0) \hPhi_\mm{dr}(t_0) \hS_\mm{dr}^{-1}(0).
    \label{eq:SupAdDrFlowRelation}
\end{equation}
Following the general derivation under ``Boundary conditions for the dressing transformation'' in the Methods,
$\hS_\mm{dr}^{-1}(0)$ transforms the initial state into the dressed frame, $\hPhi_\mm{dr}(t_0)$ describes the dressed-frame
flow, and $\hS_\mm{dr}(t_0)$ transforms the final state back to the instantaneous-eigenstate frame. Because the dressing
transformation has the same value at both boundaries, the final transformation reverses the initial change of frame whenever the
common boundary transformation commutes with the relevant dressed-frame endpoint flow. For the present two-mode construction, the
dressing transformation at the boundaries is diagonal in the instantaneous-eigenstate frame and does not mix the two transported
branches. It therefore leaves the desired endpoint map unchanged even though the dressing transformation does not reduce to the
identity.

Through the algebraic control relations, the boundary values in Eq.~\eqref{eq:SupDressingAngleBoundaryValues} also give
\begin{equation}
    \begin{aligned}
        g_x(0) &= g_x(t_0) = 0, \\
        g_z(0) &= g_z(t_0) = 0.
    \end{aligned}
    \label{eq:SupControlBoundaryConditions}
\end{equation}
The modified and bare Hamiltonians consequently coincide at the beginning and end of the protocol and share the same endpoint
instantaneous eigenstates. This physical matching of the endpoint Hamiltonians should be distinguished from the boundary
conditions on the auxiliary dressing transformation. The vanishing physical controls ensure that preparation and readout refer to
the instantaneous eigenstates of the original exceptional-point loop, whereas the equal nontrivial dressing transformations ensure
that the final change of frame reverses the initial one without altering the desired endpoint flow.

\subsection*{Full-cycle flow constraint}

In the two-mode construction, local cancellation of one dressed-frame transition channel leaves a single residual transition
direction. This is a simplifying feature of the minimal two-mode model. As a result, the full-cycle residual-flow condition
reduces to a scalar constraint on the accumulated residual interaction-picture coupling.

The full-cycle flow constraint is best expressed in the interaction-picture frame defined by the free flow $\hat{\Phi}_0
(t)$ generated by the diagonal part of Eq.~\eqref{eq:HModDr2Mode}
\begin{equation}
    \hH_0 (t) = Z(t) \hat{\sigma}_z.
    \label{eq:H02ModeIntPict}
\end{equation}
Since $\hH_0 (t)$ is a diagonal operator, we have
\begin{equation}
    \hat{\Phi}_0 (t) = \cos\left[ \beta_z (t) \right] \mathbbm{1} - i \sin\left[ \beta_z (t) \right] \hat{\sigma}_z,
    \label{eq:H02Mode}
\end{equation}
which leads to the residual interaction-picture Hamiltonian
\begin{equation}
    \hV_\uI (t) = \hat{\Phi}_0^{-1} (t) \hV_{\bar{l}} (t) \hat{\Phi}_0 (t)  = \tilde{\beta}_- (t) \hsigma_-,
    \label{eq:VI2ModeSup}
\end{equation}
with
\begin{equation}
    \label{eq:betaMinusIntPict}
    \begin{aligned}
        \tilde{\beta}_-(t)&=\frac{1}{4}e^{-i\left[2\int \alpha(t)\,dt+\mu_z(t)\right]}\left\{-2i\,\mm{R}_z(t)\left[\cos\frac{\mu_x(t)}{2}\right]^2+\left[1+\cos\mu_x(t)\right]\left[\mm{Im}[\dot\theta(t)]-i\mm{Re}[\dot\theta(t)]\right]\right.\\
        &\phantom{={}} +2e^{2i\mu_z(t)}\left[\sin\frac{\mu_x(t)}{2}\right]^2\left[-\mm{Im}[\dot\theta(t)]+i\mm{Re}[\dot\theta(t)]-i\mm{R}_z(t)-\mm{R}_x(t)\right]\\
        &\phantom{={}} \left.+2ie^{i\mu_z(t)}\left[\vphantom{2i\left[\mm{Im}[\lambda(t)]-\frac{1}{2}\tanh\{\mm{Im}[\theta(t)]\}\mm{R}_x(t)\right]}\dot{\mu}_x(t)+\sin\mu_x(t)\left(\vphantom{2i\left[\mm{Im}[\lambda(t)]-\frac{1}{2}\tanh\{\mm{Im}[\theta(t)]\}\mm{R}_x(t)\right]}2\mm{Re}[\lambda(t)]-\dot{\mu}_z(t)+\coth\{\mm{Im}[\theta(t)]\}\mm{R}_z(t)\right.\right.\right.\\
        &\phantom{={}} \left.\vphantom{\left[\cos\frac{\mu_x(t)}{2}\right]^2}\left.\left.+2i\left[\mm{Im}[\lambda(t)]-\frac{1}{2}\tanh\{\mm{Im}[\theta(t)]\}\mm{R}_x(t)\right]\right)\right]-\left[1+\cos\mu_x(t)\right]\mm{R}_x(t)\right\},
    \end{aligned}
\end{equation}
where
\begin{equation}
    \begin{aligned}
    \mm{R}_x(t)&=\frac{\mm{A}_x(t)}{\mm{B}_x(t)},\\
    \mm{R}_z(t)&=\frac{\mm{A}_z(t)}{\mm{B}_z(t)},
    \label{eq:RxRzDefinitions}
    \end{aligned}
\end{equation}
which uses
\begin{equation}
    \begin{aligned}
        \mm{A}_x(t)&=\left\{\sin\mu_x(t)+\cos\mu_z(t)\tanh\{\mm{Im}[\theta(t)]\}\right\}\left[2\mm{Im}[\lambda(t)]\sin\mu_x(t)-\mm{Im}[\dot\theta(t)]\cos\mu_x(t)\cos\mu_z(t)\right.\\
        &\phantom{={}} \left.-\mm{Re}[\dot\theta(t)]\sin\mu_z(t)-\dot{\mu}_x(t)\vphantom{\left[2\mm{Im}[\lambda(t)]\sin\mu_x(t)-\mm{Im}[\dot\theta(t)]\cos\mu_x(t)\cos\mu_z(t)\right.}\right]+\cos\mu_x(t)\sin\mu_z(t)\tanh\{\mm{Im}[\theta(t)]\}\\
        &\phantom{={}} \times\left\{-\mm{Im}[\dot\theta(t)]\sin\mu_z(t)+\mm{Re}[\dot\theta(t)]\cos\mu_x(t)\cos\mu_z(t)+\sin\mu_x(t)\left(\dot{\mu}_z(t)-2\mm{Re}[\lambda(t)]\right)\right\},\\
        \mm{B}_x(t)&=\cos^2\mu_x(t)\tanh\{\mm{Im}[\theta(t)]\}\sin^2\mu_z(t)\\
        &\phantom{={}} +\left(\sin\mu_x(t)+\cos\mu_z(t)\tanh\{\mm{Im}[\theta(t)]\}\right)\left(\cos\mu_z(t)+\sin\mu_x(t)\tanh\{\mm{Im}[\theta(t)]\}\right),
        \label{eq:AxBxBlock}
    \end{aligned}
\end{equation}
and
\begin{equation}
    \begin{aligned}
        \mm{A}_z(t)&=\coth\{\mm{Im}[\theta(t)]\}\sec\mu_x(t)\mm{Re}[\dot\theta(t)]\cot^2\mu_z(t)-\coth\{\mm{Im}[\theta(t)]\}\sec^2\mu_x(t)\mm{Im}[\dot\theta(t)]\cot\mu_z(t)\\
        &\phantom{={}} +\coth\{\mm{Im}[\theta(t)]\}\mm{Im}[\dot\theta(t)]\cot\mu_z(t)+\csc\mu_z(t)\tan\mu_x(t)\mm{Re}[\dot\theta(t)]\cot\mu_z(t)\\
        &\phantom{={}} -2\coth\{\mm{Im}[\theta(t)]\}\csc\mu_z(t)\tan\mu_x(t)\mm{Im}[\lambda(t)]+\coth\{\mm{Im}[\theta(t)]\}\csc\mu_z(t)\sec\mu_x(t)\dot{\mu}_x(t)\\
        &\phantom{={}} -2\csc\mu_z(t)\tan\mu_x(t)\left[\cot\mu_z(t)\coth\{\mm{Im}[\theta(t)]\}\sec\mu_x(t)+\csc\mu_z(t)\tan\mu_x(t)\right]\mm{Re}[\lambda(t)]\\
        &\phantom{={}} -\csc\mu_z(t)\sec\mu_x(t)\tan\mu_x(t)\mm{Im}[\dot\theta(t)]+\coth\{\mm{Im}[\theta(t)]\}\sec\mu_x(t)\mm{Re}[\dot\theta(t)]\\
        &\phantom{={}} +\csc\mu_z(t)\tan\mu_x(t)\left[\cot\mu_z(t)\coth\{\mm{Im}[\theta(t)]\}\sec\mu_x(t)+\csc\mu_z(t)\tan\mu_x(t)\right]\dot{\mu}_z(t),\\
        \mm{B}_z(t)&=\cot\mu_z(t)\left[\coth^2\{\mm{Im}[\theta(t)]\}+1\right]\csc\mu_z(t)\sec\mu_x(t)\tan\mu_x(t)\\
        &\phantom{={}} +\coth\{\mm{Im}[\theta(t)]\}\left[\cot^2\mu_z(t)\sec^2\mu_x(t)+1\right]+\coth\{\mm{Im}[\theta(t)]\}\csc^2\mu_z(t)\tan^2\mu_x(t).
        \label{eq:AzBlock}
    \end{aligned}
\end{equation}
Additionally,
\begin{equation}
\begin{aligned}
\alpha(t)&=\frac{1}{2}\left(\vphantom{\dot\theta}-\sin\mu_x(t)\sin\mu_z(t)\mm{L}_x(t)+\cos\mu_x(t)\left[-i\tanh\{\mm{Im}[\theta(t)]\}\mm{L}_x(t)+2\lambda(t)-\dot{\mu}_z(t)\right]\right.\\
&\phantom{={}} \left.+\coth\{\mm{Im}[\theta(t)]\}\left[\cos\mu_x(t)-i\tanh\{\mm{Im}[\theta(t)]\}\sin\mu_x(t)\sin\mu_z(t)\right]
\mm{L}_z(t)\right.\\
&\phantom{={}} \left.+\sin\mu_x(t)\cos\mu_z(t)\dot\theta(t)\vphantom{\left(-\sin\mu_x(t)\sin\mu_z(t)\mm{L}_x(t)+\cos\mu_x(t)\left[-i\tanh\{\mm{Im}[\theta(t)]\}\mm{L}_x(t)+2\lambda(t)-\dot{\mu}_z(t)\right]\right.}\right).
\end{aligned}
\label{eq:alphaCompact}
\end{equation}
where
\begin{equation}
\begin{aligned}
\mm{L}_x(t)&=\frac{\mm{C}_x(t)}{\mm{D}_x(t)},\\
\mm{L}_z(t)&=\frac{\mm{C}_z(t)}{\mm{D}_z(t)}.
\end{aligned}
\label{eq:HxHzAlphaDefinitions}
\end{equation}
such that
\begin{equation}
\begin{aligned}
\mm{C}_x(t)&=\left[\tanh\{\mm{Im}[\theta(t)]\}\cos\mu_z(t)+\sin\mu_x(t)\right]\left[-\mm{Im}[\dot\theta(t)]\cos\mu_x(t)\cos\mu_z(t)\right.\\
&\phantom{={}} \left.-\mm{Re}[\dot\theta(t)]\sin\mu_z(t)+2\mm{Im}[\lambda(t)]\sin\mu_x(t)-\dot{\mu}_x(t)\vphantom{\left[-\mm{Im}[\dot\theta(t)]\cos\mu_x(t)\cos\mu_z(t)\right.}\right]+\tanh\{\mm{Im}[\theta(t)]\}\cos\mu_x(t)\sin\mu_z(t)\\
&\phantom{={}} \times\left\{-\mm{Im}[\dot\theta(t)]\sin\mu_z(t)+\mm{Re}[\dot\theta(t)]\cos\mu_x(t)\cos\mu_z(t)+\sin\mu_x(t)\left(\dot{\mu}_z(t)-2\mm{Re}[\lambda(t)]\right)\right\},\\
\mm{D}_x(t)&=\tanh\{\mm{Im}[\theta(t)]\}\cos^2\mu_x(t)\sin^2\mu_z(t)\\
&\phantom{={}} +\left[\tanh\{\mm{Im}[\theta(t)]\}\cos\mu_z(t)+\sin\mu_x(t)\right]\left[\tanh\{\mm{Im}[\theta(t)]\}\sin\mu_x(t)+\cos\mu_z(t)\right],
\label{eq:CxDxAlpha}
\end{aligned}
\end{equation}
\begin{equation}
\begin{aligned}
\mm{C}_z(t)&=-\mm{Im}[\dot\theta(t)]\coth\{\mm{Im}[\theta(t)]\}\sec^2\mu_x(t)\cot\mu_z(t)-\mm{Im}[\dot\theta(t)]\tan\mu_x(t)\sec\mu_x(t)\csc\mu_z(t)\\
&\phantom{={}} +\mm{Im}[\dot\theta(t)]\coth\{\mm{Im}[\theta(t)]\}\cot\mu_z(t)+\coth\{\mm{Im}[\theta(t)]\}\mm{Re}[\dot\theta(t)]\sec\mu_x(t)\cot^2\mu_z(t)\\
&\phantom{={}} +\coth\{\mm{Im}[\theta(t)]\}\mm{Re}[\dot\theta(t)]\sec\mu_x(t)-2\mm{Im}[\lambda(t)]\coth\{\mm{Im}[\theta(t)]\}\tan\mu_x(t)\csc\mu_z(t)\\
&\phantom{={}} -2\mm{Re}[\lambda(t)]\tan\mu_x(t)\csc\mu_z(t)\left[\coth\{\mm{Im}[\theta(t)]\}\sec\mu_x(t)\cot\mu_z(t)+\tan\mu_x(t)\csc\mu_z(t)\right]\\
&\phantom{={}} +\coth\{\mm{Im}[\theta(t)]\}\dot{\mu}_x(t)\sec\mu_x(t)\csc\mu_z(t)+\mm{Re}[\dot\theta(t)]\tan\mu_x(t)\cot\mu_z(t)\csc\mu_z(t)\\
&\phantom{={}} +\tan\mu_x(t)\dot{\mu}_z(t)\csc\mu_z(t)\left[\coth\{\mm{Im}[\theta(t)]\}\sec\mu_x(t)\cot\mu_z(t)+\tan\mu_x(t)\csc\mu_z(t)\right],\\
\mm{D}_z(t)&=\coth\{\mm{Im}[\theta(t)]\}\left[\sec^2\mu_x(t)\cot^2\mu_z(t)+1\right]+\coth\{\mm{Im}[\theta(t)]\}\tan^2\mu_x(t)\csc^2\mu_z(t)\\
&\phantom{={}} +\left[\coth^2\{\mm{Im}[\theta(t)]\}+1\right]\tan\mu_x(t)\sec\mu_x(t)\cot\mu_z(t)\csc\mu_z(t).
\end{aligned}
\label{eq:CzDzAlpha}
\end{equation}

Since $\hV_\uI (t)$ commutes with itself at all times, the requirement that the residual interaction-picture flow
have no net effect over the cycle can be expressed as
\begin{equation}
    \hat{\Phi}_\uI (t_0) = \exp\left[ -i \int_0^{t_0} \di{t} \tilde{\beta} (t) \hsigma_-  \right] = \mathbbm{1}.
    \label{eq:AvgCondIntPictFlow}
\end{equation}
This is equivalent to requiring that
\begin{equation}
    \int_0^{t_0} \di{t} \tilde{\beta}_-(t) = 0.
    \label{eq:AvgCondIntPictFlowCoeff}
\end{equation}

To satisfy Eq.~\eqref{eq:AvgCondIntPictFlowCoeff}, we determine the Fourier coefficients defining $\mu_x(t)$ and $\mu_z(t)$ [see
Eq.~\eqref{eq:DressingAnglesOptSup}] by minimizing the squared residual
\begin{equation}
    \abs{\int_0^{t_0} \di{t} \tilde{\beta}_-(t)}^2,
    \label{eq:AvgCondIntPictFlowCoeffOpt}
\end{equation}
and restricting the search to the intervals $-5 \leq a_{j,l}, b_{j,m} \leq 5$ with $l=m=3$ and $j \in \{x,z\}$. This minimization
is used as a stable root-finding method for the full-cycle constraint, rather than as an optimization of the state-transfer error.

\suppnote{Projected-subspace version of the full-cycle flow constraint}
\label{SupSec:RelaxedFlowConstraint}

This Supplementary Note generalizes the full-cycle flow constraint to protocols in which only a selected subspace of
holomorphically continued instantaneous eigenstates is required to undergo the target transport. In this case, the residual
interaction-picture flow need not be trivial on the full Hilbert space.

Let $\hP$ project onto the target subspace and let $\hQ=\mathbbm{1}-\hP$ project onto its complement. The residual
interaction-picture flow must reproduce the identity within the target subspace and must not couple that subspace to its
complement at the final time. A sufficient set of conditions is
\begin{equation}
    \begin{aligned}
        \hP \hPhi_\uI (t_0) \hP &= \hP, \\
        \hP \hPhi_\uI (t_0) \hQ &= \mathbf{0}, \\
        \hQ \hPhi_\uI (t_0) \hP &= \mathbf{0}.
    \end{aligned}
    \label{eq:SupProjectedFlowConstraint}
\end{equation}
The first line removes the accumulated residual dynamics within the target subspace, while the remaining lines prevent final-time
leakage between that subspace and its complement. No condition is imposed on $\hQ\hPhi_\uI (t_0)\hQ$, because the flow within the
complementary subspace is irrelevant to the target operation.

The same boundary conditions derived under ``Boundary conditions for the dressing transformation'' in the Methods apply to the
selected subspace. In particular, the initial and final changes of frame must preserve the target subspace and must not alter the
desired endpoint flow within that subspace.  These boundary conditions are distinct from the projected full-cycle constraint
above, which requires the accumulated residual flow to be trivial within the target subspace and prevents final-time coupling
between that subspace and its complement. Restricting the construction in this way reduces the number of full-cycle constraints
when only a subset of states is relevant.

\suppnote{Alternative locally cancelled channel in the two-mode model}
\label{SupSec:TwoModeAlternate}

This Supplementary Note outlines the same prescription given in \suppref{SupSec:TwoModeLocalGlobal}, but swaps the choice of
locally cancelled dressed-frame coupling channel and residual coupling channel. Here, we choose to cancel the coefficient
$\beta_-(t)$ [see Eq.~\eqref{eq:CoeffsHModDr2Mode}] locally. The residual interaction-picture coupling $\tilde{\beta}_+(t)$ then
enters the full-cycle flow constraint for the remaining channel. We have
\begin{equation}
    \begin{aligned}
        \tilde{\beta}_+(t)&=\frac{1}{4}e^{2i\int \zeta(t)\,dt-i\mu_z(t)}\left\{-2ie^{i\mu_z(t)}\left[\mm{K}_z(t)-i\dot\theta(t)\left[\sin\mu_z(t)-i\cos\mu_x(t)\cos\mu_z(t)\right]\right.\right.\\
        &\phantom{={}} \left.\left.+\sin\mu_x(t)\left[2\lambda(t)-\dot{\mu}_z(t)\right]-i\dot{\mu}_x(t)\vphantom{\left[\mm{K}_z(t)-i\dot\theta(t)\left[\sin\mu_z(t)-i\cos\mu_x(t)\cos\mu_z(t)\right]\right.}\right]\right.\\
    &\phantom{={}} \left.-\left[2e^{i\mu_z(t)}\tanh\{\mm{Im}[\theta(t)]\}\sin\mu_x(t)+e^{2i\mu_z(t)}\left[\cos\mu_x(t)+1\right]-\cos\mu_x(t)+1\right]\mm{K}_x(t)\vphantom{\left\{-2ie^{i\mu_z(t)}\left[\mm{K}_z(t)-i\dot\theta(t)\left[\sin\mu_z(t)-i\cos\mu_x(t)\cos\mu_z(t)\right]\right.\right.}\right\}.
    \end{aligned}
\label{eq:tildeBetaPlusCompact}
\end{equation}
where
\begin{equation}
\begin{aligned}
\mm{K}_x(t)&=\frac{\mm{P}_x(t)}{\mm{Q}_x(t)},\\\mm{K}_z(t)
&=\frac{2\coth\{\mm{Im}[\theta(t)]\}\left\{\sin\mu_x(t)+\tanh\{\mm{Im}[\theta(t)]\}\left[\cos\mu_z(t)+i\cos\mu_x(t)\sin\mu_z(t)\right]\right\}\mm{P}_z(t)}{\mm{Q}_z(t)},
\end{aligned}
\label{eq:KxKzDefinitions}
\end{equation}
which is comprised of
\begin{equation}
\begin{aligned}
\mm{P}_x(t)&=\left[\sin\mu_x(t)-\tanh\{\mm{Im}[\theta(t)]\}\cos\mu_z(t)\right]\left[-\mm{Im}[\dot\theta(t)]\cos\mu_x(t)\cos\mu_z(t)\right.\\
&\phantom{={}} \left.+\mm{Re}[\dot\theta(t)]\sin\mu_z(t)+2\mm{Im}[\lambda(t)]\sin\mu_x(t)+\dot{\mu}_x(t)\vphantom{\left[-\mm{Im}[\dot\theta(t)]\cos\mu_x(t)\cos\mu_z(t)\right.}\right]+\tanh\{\mm{Im}[\theta(t)]\}\cos\mu_x(t)\sin\mu_z(t)\\
&\phantom{={}} \times\left[\mm{Im}[\dot\theta(t)]\sin\mu_z(t)+\mm{Re}[\dot\theta(t)]\cos\mu_x(t)\cos\mu_z(t)+\sin\mu_x(t)\left(\dot{\mu}_z(t)-2\mm{Re}[\lambda(t)]\right)\right],\\\mm{Q}_x(t)
&=\tanh\{\mm{Im}[\theta(t)]\}\cos^2\mu_x(t)\sin^2\mu_z(t)\\
&\phantom{={}} +\left[\tanh\{\mm{Im}[\theta(t)]\}\cos\mu_z(t)-\sin\mu_x(t)\right]\left[\cos\mu_z(t)-\tanh\{\mm{Im}[\theta(t)]\}\sin\mu_x(t)\right],
\end{aligned}
\label{eq:PxQxTildeBetaPlus}
\end{equation}
and
\begin{equation}
\begin{aligned}
\mm{P}_z(t)&=\mm{Im}[\dot\theta(t)]\coth\{\mm{Im}[\theta(t)]\}\sec^2\mu_x(t)\cot\mu_z(t)-\mm{Im}[\dot\theta(t)]\tan\mu_x(t)\sec\mu_x(t)\csc\mu_z(t)\\
&\phantom{={}} -\mm{Im}[\dot\theta(t)]\coth\{\mm{Im}[\theta(t)]\}\cot\mu_z(t)+\coth\{\mm{Im}[\theta(t)]\}\mm{Re}[\dot\theta(t)]\sec\mu_x(t)\cot^2\mu_z(t)\\
&\phantom{={}} +\coth\{\mm{Im}[\theta(t)]\}\mm{Re}[\dot\theta(t)]\sec\mu_x(t)+2\mm{Im}[\lambda(t)]\coth\{\mm{Im}[\theta(t)]\}\tan\mu_x(t)\csc\mu_z(t)\\
&\phantom{={}} +2\mm{Re}[\lambda(t)]\tan\mu_x(t)\csc\mu_z(t)\left[\tan\mu_x(t)\csc\mu_z(t)-\coth\{\mm{Im}[\theta(t)]\}\sec\mu_x(t)\cot\mu_z(t)\right]\\
&\phantom{={}} +\coth\{\mm{Im}[\theta(t)]\}\dot{\mu}_x(t)\sec\mu_x(t)\csc\mu_z(t)+\tan\mu_x(t)\dot{\mu}_z(t)\csc\mu_z(t)\\
&\phantom{={}} \times\left[\coth\{\mm{Im}[\theta(t)]\}\sec\mu_x(t)\cot\mu_z(t)-\tan\mu_x(t)\csc\mu_z(t)\right]\\
&\phantom{={}}-\mm{Re}[\dot\theta(t)]\tan\mu_x(t)\cot\mu_z(t)\csc\mu_z(t),\\
\mm{Q}_z(t)&=-2\coth\{\mm{Im}[\theta(t)]\}\left[\sec^2\mu_x(t)\cot^2\mu_z(t)+1\right]-2\coth\{\mm{Im}[\theta(t)]\}\tan^2\mu_x(t)\csc^2\mu_z(t)\\
&\phantom{={}} +2\left[\coth^2\{\mm{Im}[\theta(t)]\}+1\right]\tan\mu_x(t)\sec\mu_x(t)\cot\mu_z(t)\csc\mu_z(t).
\end{aligned}
\label{eq:PzQzTildeBetaPlus}
\end{equation}
Lastly,
\begin{equation}
    \label{eq:zetaCompact}
    \begin{aligned}
        \zeta(t)&=\frac{1}{2}\left\{\vphantom{\dot\theta}-\sin\mu_x(t)\sin\mu_z(t)\mm{L}_x(t)+\cos\mu_x(t)\left[-i\tanh\{\mm{Im}[\theta(t)]\}\mm{L}_x(t)+2\lambda(t)-\dot{\mu}_z(t)\right]\right.\\
        &\phantom{={}} \left.+\coth\{\mm{Im}[\theta(t)]\}\left[\cos\mu_x(t)-i\tanh\{\mm{Im}[\theta(t)]\}\sin\mu_x(t)\sin\mu_z(t)\right]\mm{L}_z(t)\right.\\
        &\phantom{={}} \left.+\sin\mu_x(t)\cos\mu_z(t)\dot\theta(t)\right\}.
    \end{aligned}
\end{equation}
where
\begin{equation}
\begin{aligned}
\mm{L}_x(t)&=\frac{\mm{E}_x(t)}{\mm{F}_x(t)}\\
\mm{L}_z(t)&=\frac{\mm{E}_z(t)}{\mm{F}_z(t)}.
\end{aligned}
\label{eq:LxLzZetaDefinitions}
\end{equation}
in which
\begin{equation}
\begin{aligned}
\mm{E}_x(t)&=\left[\tanh\{\mm{Im}[\theta(t)]\}\cos\mu_z(t)+\sin\mu_x(t)\right]\left[-\mm{Im}[\dot\theta(t)]\cos\mu_x(t)\cos\mu_z(t)\right.\\
&\phantom{={}} \left.-\mm{Re}[\dot\theta(t)]\sin\mu_z(t)+2\mm{Im}[\lambda(t)]\sin\mu_x(t)-\dot{\mu}_x(t)\vphantom{\left[-\mm{Im}[\dot\theta(t)]\cos\mu_x(t)\cos\mu_z(t)\right.}\right]+\tanh\{\mm{Im}[\theta(t)]\}\cos\mu_x(t)\sin\mu_z(t)\\
&\phantom{={}} \left\{-\mm{Im}[\dot\theta(t)]\sin\mu_z(t)+\mm{Re}[\dot\theta(t)]\cos\mu_x(t)\cos\mu_z(t)+\sin\mu_x(t)\left[\dot{\mu}_z(t)-2\mm{Re}[\lambda(t)]\right]\right\},\\
\mm{F}_x(t)&=\tanh\{\mm{Im}[\theta(t)]\}\cos^2\mu_x(t)\sin^2\mu_z(t)\\
&\phantom{={}} +\left[\tanh\{\mm{Im}[\theta(t)]\}\cos\mu_z(t)+\sin\mu_x(t)\right]\left[\tanh\{\mm{Im}[\theta(t)]\}\sin\mu_x(t)+\cos\mu_z(t)\right],
\end{aligned}
\label{eq:ExFxZeta}
\end{equation}
and
\begin{equation}
\begin{aligned}
\mm{E}_z(t)&=-\mm{Im}[\dot\theta(t)]\coth\{\mm{Im}[\theta(t)]\}\sec^2\mu_x(t)\cot\mu_z(t)-
\mm{Im}[\dot\theta(t)]\tan\mu_x(t)\sec\mu_x(t)\csc\mu_z(t)\\
&\phantom{={}} +\mm{Im}[\dot\theta(t)]\coth\{\mm{Im}[\theta(t)]\}\cot\mu_z(t)+\coth\{\mm{Im}[\theta(t)]\}\mm{Re}[\dot\theta(t)]\sec\mu_x(t)\cot^2\mu_z(t)\\
&\phantom{={}} +\coth\{\mm{Im}[\theta(t)]\}\mm{Re}[\dot\theta(t)]\sec\mu_x(t)-2\mm{Im}[\lambda(t)]\coth\{\mm{Im}[\theta(t)]\}\tan\mu_x(t)\csc\mu_z(t)\\
&\phantom{={}} -2\mm{Re}[\lambda(t)]\tan\mu_x(t)\csc\mu_z(t)\left[\coth\{\mm{Im}[\theta(t)]\}\sec\mu_x(t)\cot\mu_z(t)+\tan\mu_x(t)\csc\mu_z(t)\right]\\
&\phantom{={}} +\coth\{\mm{Im}[\theta(t)]\}\dot{\mu}_x(t)\sec\mu_x(t)\csc\mu_z(t)+\mm{Re}[\dot\theta(t)]\tan\mu_x(t)\cot\mu_z(t)\csc\mu_z(t)\\
&\phantom{={}} +\tan\mu_x(t)\dot{\mu}_z(t)\csc\mu_z(t)\left[\coth\{\mm{Im}[\theta(t)]\}\sec\mu_x(t)\cot\mu_z(t)+\tan\mu_x(t)\csc\mu_z(t)\right],\\
\mm{F}_z(t)&=\coth\{\mm{Im}[\theta(t)]\}\left[\sec^2\mu_x(t)\cot^2\mu_z(t)+1\right]+\coth\{\mm{Im}[\theta(t)]\}\tan^2\mu_x(t)\csc^2\mu_z(t)\\
&\phantom{={}} +\left[\coth^2\{\mm{Im}[\theta(t)]\}+1\right]\tan\mu_x(t)\sec\mu_x(t)\cot\mu_z(t)\csc\mu_z(t).
\end{aligned}
\label{eq:EzFzZeta}
\end{equation}
\begin{figure}[t]
    \centering
    \includegraphics[scale=0.5]{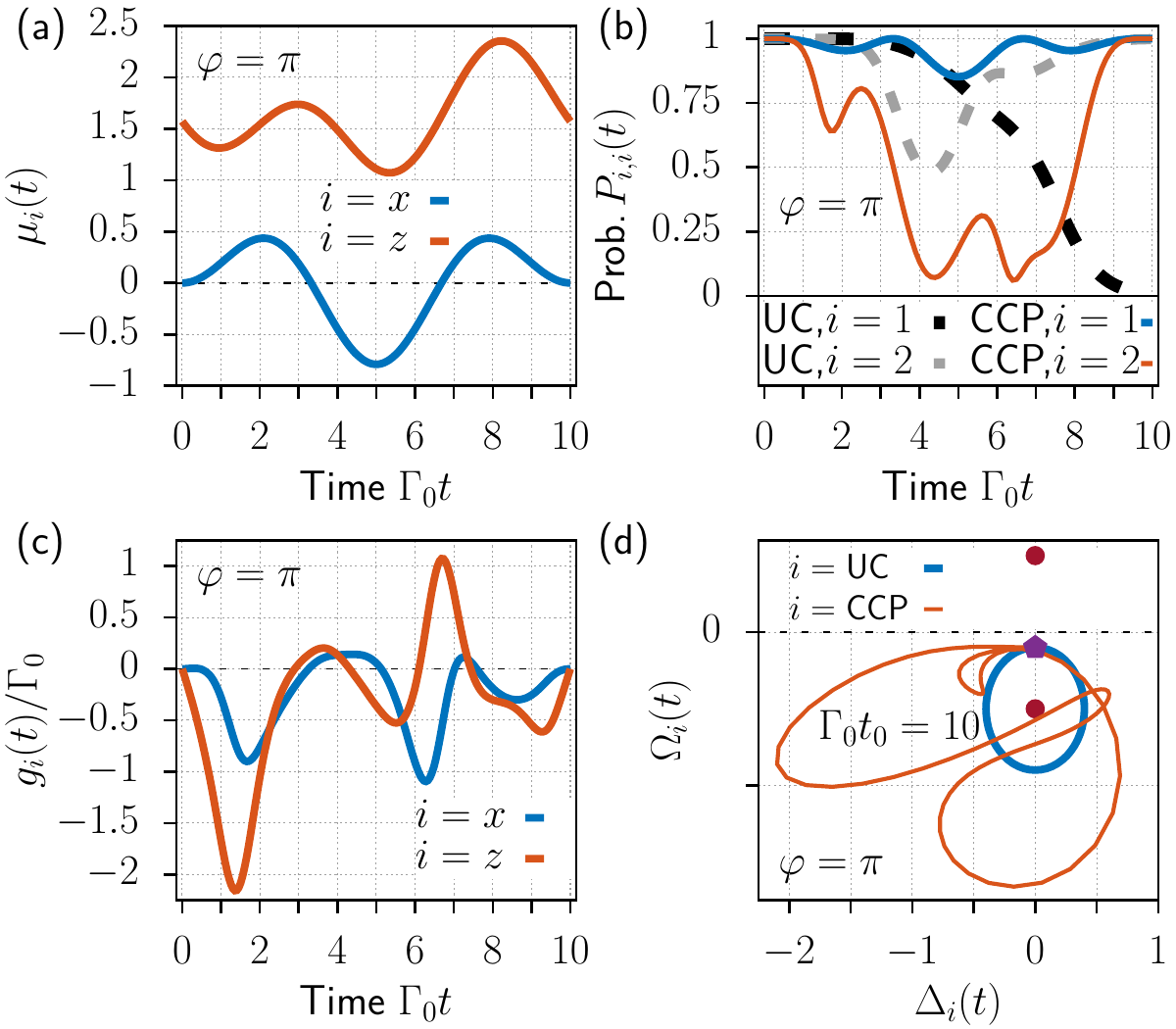}
    \caption{
        \textbf{Coherent controls for a representative loop}. (a) Dressing angles $\mu_x (t)$ and$\mu_z (t)$. (b)
        Branch-transition probabilities in the holomorphically continued instantaneous-eigenstate frame for uncontrolled and
        constrained-control dynamics.  (c) Native laboratory-frame controls $g_x(t)$ and $g_z(t)$, which vanish at the protocol
        boundaries. (d) Uncorrected and coherent controlled trajectories in the$(\Delta,\Omega)$ plane.
    }
    \label{fig:OptimizationResults}
\end{figure}
The alternative choice of locally cancelled channel uses the same boundary conditions as the construction in
\ref{SupSec:TwoModeLocalGlobal}. In particular, the dressing transformation has the same nontrivial value at the beginning and
end of the protocol, so the final change of frame reverses the initial one without altering the desired endpoint map. The physical
control corrections also satisfy $g_x(0)=g_x(t_0)=g_z(0)=g_z(t_0)=0$, ensuring that the bare and modified Hamiltonians coincide at
both boundaries. These properties are independent of which of the two dressed-frame coupling channels is cancelled locally.

Supplementary Fig.~\ref{fig:OptimizationResults} shows a representative protocol obtained by cancelling the opposite dressed-frame
coupling channel from the one cancelled in the main text. The resulting smooth waveforms within the native laboratory control
manifold reproduce the endpoint state transfer selected by the exceptional-point loop while satisfying the same boundary
conditions as the main construction. This confirms that the method is not tied to a particular choice of locally cancelled
coupling channel.

Supplementary Fig.~\ref{fig:OptimizationResults}a shows the dressing angles $\mu_x(t)$ and $\mu_z(t)$ obtained from Fourier
coefficients satisfying the full-cycle flow constraint. Supplementary Fig.~\ref{fig:OptimizationResults}b compares the uncorrected
dynamics with the constrained-control dynamics in the holomorphically continued instantaneous-eigenstate frame. The uncorrected
dynamics does not reproduce the target final state transfer, whereas the constrained-control protocol maps both initial states to
their corresponding target final states. Supplementary Fig.~\ref{fig:OptimizationResults}c shows the resulting real
laboratory-frame fields $g_x(t)$ and $g_z(t)$, which vanish at both protocol boundaries. The bare and modified Hamiltonians
therefore share the same initial and final instantaneous eigenstates for this alternative local-cancellation choice. Supplementary
Fig.~\ref{fig:OptimizationResults}d shows the associated trajectory in the $(\Delta,\Omega)$ plane.

\end{document}